\documentclass[fleqn,usenatbib]{mnras}

\usepackage{newtxtext,newtxmath}

\usepackage[T1]{fontenc}

\usepackage{graphicx}	
\usepackage{amsmath,comment}	
\usepackage{natbib}

\title[O-IR QPOs in Swift J1727.8--1613]
{Sub-second optical/near-infrared quasi-periodic oscillations from the black hole X-ray transient Swift~J1727.8--1613}

\author[F. M. Vincentelli et al. ]{F. M. Vincentelli$^{1,2,3,4}$,\thanks{E-mail: vincentelli.astro@gmail.com}
T. Shahbaz$^{2,3}$, 
P. Casella$^{4}$, 
 V. S. Dhillon$^{5,2}$, 
 J. Paice${^6}$, D. Altamirano$^{1}$, 
 \newauthor
 N. Castro Segura${^7}$, R.  Fender${^8}$, P. Gandhi$^{1}$, S. Littlefair${^5}$, T.  Maccarone${^9}$, J. Malzac${^{10}}$, K. O'Brien${^6}$
 \newauthor
 D. M. Russell${^{11}}$, A. J.  Tetarenko${^{12}}$, P. Uttley${^{13}}$, A. Veledina${^{14,15}}$
\\
$^{1}$Department of Physics and Astronomy, University of Southampton, Highfield, Southampton, SO17 1BJ, UK \\
$^{2}$Instituto de Astrof\'\i{}sica de Canarias (IAC), E-38200 La Laguna, 
Tenerife, Spain \\
$^{3}$Departamento de  Astrof\'\i{}sica, Universidad de La Laguna (ULL), 
E-38206 La Laguna, Tenerife, Spain \\
$^{4}$INAF-Osservatorio Astronomico di Roma, Via Frascati 33, I-00076, Monte Porzio Catone (RM), Italy \\
$^{5}$ Astrophysics Research Cluster, School of Mathematical and Physical Sciences, University of Sheffield, Sheffield, S3 7RH, UK\\
$^6$Centre for Advanced Instrumentation, Department of Physics, Durham University, Durham, UK\\
$^7$Department of Physics, Gibbet Hill Road, University of Warwick, Coventry, CV4 7AL, UK\\
$^8$Astrophysics, Department of Physics, University of Oxford, Keble Road, Oxford OX1 3RH, UK\\
$^9$Department of Physics \& Astronomy, Texas Tech University, Box 41051, Lubbock, TX, USA, 79409-1051\\
$^{10}$ IRAP, Universit\'{e} de Toulouse, CNRS, UPS, CNES, Toulouse, France \\
$^{11}$Center for Astrophysics and Space Science (CASS), New York University Abu Dhabi, PO Box 129188, Abu Dhabi, UAE\\
$^{12}$Department of Physics and Astronomy, University of Lethbridge, Lethbridge, Alberta, T1K 3M4, Canada\\
$^{13}$Anton Pannekoek Institute for Astronomy, University of Amsterdam, Science Park 904, 1098 XH Amsterdam, The Netherlands\\
$^{14}$Department of Physics and Astronomy, FI-20014 University of Turku, Finland\\
$^{15}$ Nordita, Stockholm University and KTH Royal Institute of Technology, Hannes Alfv\'ens v\"ag 12, SE-10691 Stockholm, Sweden
}

\date{Accepted XXX. Received YYY; in original form ZZZ}

\pubyear{\the\year{}}

\begin{document}
\label{firstpage}
\pagerange{\pageref{firstpage}--\pageref{lastpage}}
\maketitle

\begin{abstract}
We report on the detection of  optical/near-infrared (O-IR) quasi-periodic oscillations (QPOs) from the black hole X-ray transient Swift\,J1727.8--1613. We obtained three X-ray and O-IR high-time-resolution observations of the source during {its intermediate state} (2023 September 9, 15 and 17) using NICER, HAWK-I@VLT, HIPERCAM@GTC and ULTRACAM@NTT. We clearly detected a QPO in the X-ray and O-IR bands during all three epochs. The QPO evolved, drifting from 1.4\,Hz in the first epoch, up to 2.2\,Hz in the second and finally reaching 4.2\,Hz at the third epoch.  These are among the highest O-IR QPO frequencies detected for a black hole X-ray transient. During the first two epochs, the X-ray and O-IR emission are correlated, with an optical lag (compared to the X-rays) varying from +70\,ms to 0\,ms.
Finally, during the third epoch, we measured for the first time, a lag of the $z_s$-band respect to the $g_s$-band at the QPO frequency ($\approx$+10\,ms).
By estimating the variable O-IR SED we find that the emission is most likely non-thermal.  Current state-of-the-art models can explain some of these properties, but neither the jet nor the hot flow model can easily explain the observed evolution of the QPOs. While this allowed us to put tight constraints on these components,  more frequent coverage of the state transition with fast multi-wavelength observations is still needed to fully understand the evolution of the disc/jet properties in BH LMXBs.
\end{abstract}




\section{Introduction} \label{sec:intro}
Black hole transients (BHT) are a class of galactic X-ray binary system which host a stellar-mass black hole accreting mass from a low-mass donor star. These systems undergo luminous outbursts with X-ray luminosities of $\approx$10$^{37-38}$ erg s$^{-1}$, which can last from a few weeks to years, displaying highly variable emission from the radio to hard X-ray bands \citep[see e.g.][]{mirabel1998,corbel2002,tetarenko2021_a}{}{}. Years of multi-wavelength studies have identified three main physical components: an optically thick, geometrically thin accretion disc, an optically thin, geometrically thick hot inflow (also referred to as a corona) and a compact jet.

The accretion disc emits thermal radiation, producing a multi-color blackbody spectrum which typically peaks in the soft X-ray band \citep[$\approx$ few keV; see e.g.][]{shakurasunayev,gierlinski2001}; the hot flow, instead, is believed to be formed by hot electrons which show a cut-off power-law (which dominates the hard X-rays) via inverse Compton scattering of lower energy seed photons \citep[see e.g.][]{haardmarasrchi,poutanen1996}. The spectral evolution of BHTs shows that the relative contribution of these components changes dramatically during the outbursts \citep{zdariskigierlinski}. Furthermore, a clear pattern was identified in the evolution of these transients, along with distinctive \lq\lq accretion states\rq\rq \citep{fender2004}. The outburst always starts in a so-called \lq\lq hard X-ray state\rq\rq, i.e. with an X-ray spectrum dominated by a non-thermal power-law; then, reached a certain luminosity, sources rapidly move towards a  \lq\lq soft X-ray state\rq\rq~  with thermal X-ray spectrum arising from an accretion disc\citep{dgk07}. 

 Compact steady jets are mostly observed in the hard state,  exhibiting a flat spectrum that spans from radio to optical-infrared (O-IR) wavelengths \citep[see e.g.][]{corbel2002,dgk07,gandhi2011}. This behavior is typically explained as the result of the superposition of synchrotron profiles arising from different populations of electrons at different energies \citep{blandford&konigl}. The overall spectrum has an optically thin branch (with a slope $\propto\nu^{-0.7}$), which flattens out for frequencies lower than the self-absorption break.  {Multi-wavelength campaigns of BH LMXBs have measured a variable break that vary throughout the outburt between $\approx$10$^{11}$ and 10$^{14}$ Hz \citep[see e.g.][]{Russell2013,Koljonen2015,russell2020,Echibur2024}.}

One of the key properties of accreting systems is the presence of strong stochastic variability from the inflow \citep[see e.g.][]{scaringi2015}{}{}. Decades of timing studies revealed that during their hard X-ray state, the X-ray variability of BHTs shows a Fourier power spectrum characterized by broadband noise extending from $\approx$10$^{-2}$ Hz up to a few tens of Hz \citep{belloni2002}. 
The amplitude and characteristic frequencies of the X-ray power spectrum evolve significantly during the outburst. As the system moves towards the soft X-ray state, the power spectrum shifts towards higher frequencies and decreases in amplitude \citep[see e.g.][]{belloni2005,heil2015}. Although this has been successfully reproduced in terms of the accretion disc gradually moving towards the last stable orbit (thus, making the hot inflow smaller), there are still many uncertainties regarding the exact geometry of the accretion flow and the physical processes involved \citep[see e.g.][]{uttley2005,uttleymalzac,rapisarda2017,veledina2018,mendez2024_hidden}. 

The development of fast optical/near-infrared (O-IR) detectors in the late 1990s--2000s   led to the discovery of previously unknown complex phenomenology \citep[see e.g.][]{eikenberry1998,kanbach2001} which opened a new way to constrain the accretion/ejection physical properties. A clear example is the detection of a $\approx$0.1s delay between the O-IR emission with respect to the X-rays \citep[][]{gandhi2008,gandhi2017,casella2010,paice2019,tetarenko2021_a}{}{}, which has been interpreted as the travel time of accretion rate fluctuations from the inflow to the base of the jet.  This measurement led to the development of the first internal-shock model for jets in X-ray binaries which reproduced the observed lag by linking the internal shells' velocity to the observed X-ray variability \citep{jamil2010,malzac2014,malzac2018}.  Further studies have also revealed the presence of a more complex relation between X-rays and O-IR emission, including, for instance, strong anti-correlations between the two bands \citep{durant2008,vincentelli2021_1535}. Although some of these features could be explained in terms of the internal shock model \citep{malzac2018}, this behavior has also been quantitatively reproduced through synchrotron radiation from the external regions of a hot magnetized flow \citep{veledina2017,ulgiati2024}. 

A characteristic feature of BHTs is the presence of quasi-periodic oscillations in the X-ray power spectrum \citep[QPOs;][]{wijnands-vanderklis1999} . These narrow components are seen almost ubiquitously across the outbursts and follow the evolution of the broadband noise. The origin of these phenomena, however, is still not fully understood. Different type of oscillations have been identified  \citep{ingram_notta2020}. { The osicllations that are most commonly found during the outburst are the so called \lq\lq type-C QPOs\rq\rq \citep{casella2005}, and they are typically found in the hard and hard-intermediate states. Intensive studies have shown that QPO, increase their frequency as these systems move towards the soft state, along with the other broadband noise components of the power-spectrum  \citep[see e.g.][]{belloni2005,ingram_notta2020}. Population studies of these oscillations have also shown that the amplitude of type-C QPOs is stronger in higher inclination sources \citep{Schnittman2006,motta2015}, suggesting a geometrical origin.} One of th  most successful models, for example, invokes Lense-Thirring precession of the hot inner flow \citep{ingram2009,ingram2016,marcel-neilsen2021,nathan2022}.  Fast photometric studies have also led to the discovery of the O-IR counterpart of X-ray QPOs \citep[]{motch1983,gandhi2010,kalamkar2016,thomas2022}{}. Similarly to the X-rays, the main models for O-IR QPOs are based on the precession of the physical component responsible for the low energy emission: i.e. the hot inflow and the jet \citep[]{veledina_poutanen_ingram_2013,malzac2018}{}. 

Due to the challenge of performing strictly simultaneous high time resolution observations of these transient systems, to date, we have only a handful of QPOs detected simultaneously at different wavelengths. Thus, it is still unclear how the X-ray and O-IR oscillations co-evolve during the outburst, or how they change with inclination. Furthermore, only one O-IR QPO has been detected for frequencies greater than $\approx$1 Hz \citep[]{vincentelli2021_1535}{}, i.e.  during the late stages of transition between the hard and the soft state \citep[also referred to as \lq\lq intermediate state\rq\rq;][]{belloni2005,munoz-darias2010,motta2011}. This limits our understanding of these oscillations, and of the inflow/outflow structure. Here, we present the result of a multi-wavelength campaign of the X-ray transient Swift\,J1727.8--1613 during such intermediate state during the 2023 outburst. At its peak, the source became one of the brightest objects in the X-ray sky in the last few years, reaching $\approx$8 Crab in X-rays \citep{palmer2023_atel} and showing strong multi-wavelength variability  with powerful jets and outflows \citep[see e.g.][]{ingram2024,matasanchez2024,wood2024}.

\section{Observations} 

\subsection{The campaign:} 
\begin{figure*}
\includegraphics[scale=0.7]{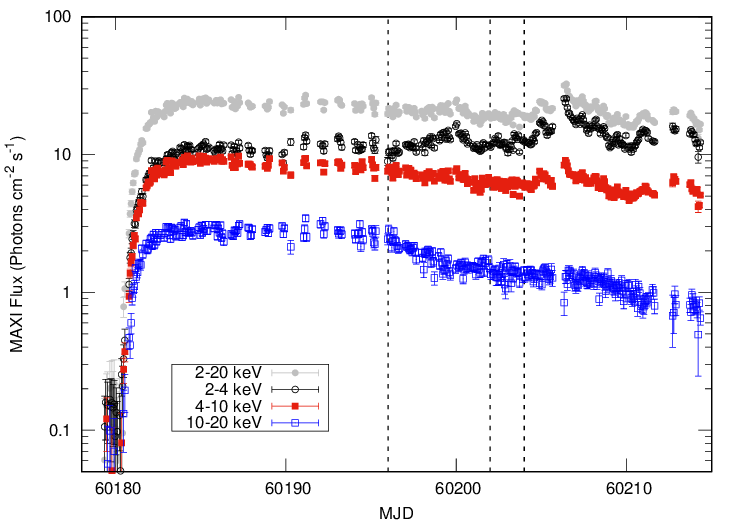}
\includegraphics[scale=0.7]{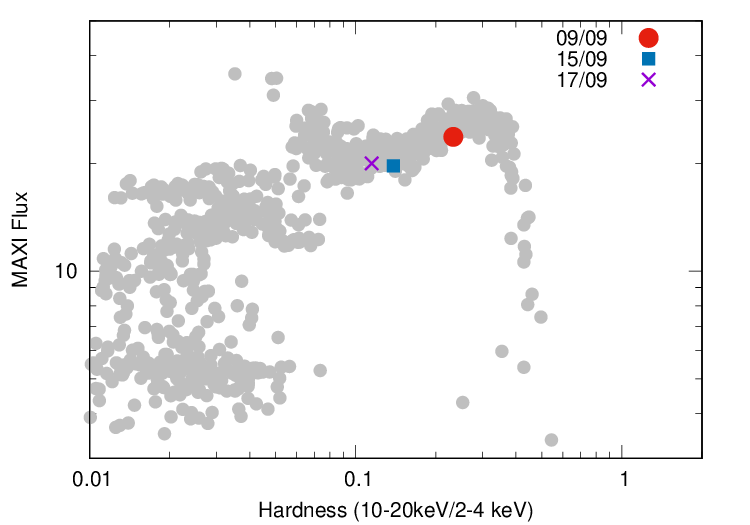}

\caption{The long-term X-ray evolution of Swift\,J1727.8--1613. \textit{Left Panel:} the MAXI daily light curve. The vertical lines show the dates of our observations. \textit{Right Panel:} the hardness-intensity diagram computed from the MAXI 2-20 keV data. The coloured points show the dates of our observations.
\label{fig:general_lc}}
\end{figure*}

We collected three high time-resolution O-IR datasets on 2023, September 9, 15, and 17. The optical data were collected with ULTRACAM@NTT and HiPERCAM@GTC. IR data were collected with HAWK-I@VLT. For the first two epochs, X-ray strictly simultaneous data were collected with NICER.  
A summary of the observations can be found in Tab. \ref{tab:log}. As shown in Fig. \ref{fig:general_lc}, the system was evolving through the intermediate state, showing a steady drop in the hard X-ray flux (10-20 keV). After extracting the time series for each band, we barycentered them using the JPL DE-430 ephemerides using the following coordinates: RA=261.930$^\circ$ and DEC=-16.205$^\circ$. All datasets were converted into Barycentric Dynamical Time (BJD\_TDB). For the optical time series, we used the method and software described in \cite{eastman_achieving_2010}. The X-ray events from NICER were corrected using the \textsc{heasoft} software \textsc{barycorr}.

\begin{table*}
\caption{Summary of our fast multi-wavelength campaign. For each date we report the instrument, band, time of observation and Observation ID (if present). }

\begin{center}

\begin{tabular}{|lllll|} 
\hline

Date       & Instrument                                            & band          & Time (UTC) & OBSID \\
\hline
\textbf{09/09/2023} &                                                       &               &      &      \\
           & HAWK-I @VLT & $K_s$   & 01:19-01:52   &   112.2615.001   \\
           & ULTRACAM@NTT                                          & $u_s$,$g_s$,$i_s$         & 23:56$^*$-03:31  &  NA    \\
           & NICER                                                 & 0.5-10 keV    & 23:53$^*$-00:01 &    6203980116  \\
           &                                                       &               &01:26-01:33 &  \\
   
\textbf{15/09/2023} &                                                       &               &      &      \\
           &HAWK-I @VLT & $K_s$  &  01:18-01:50 &    112.2615.001  \\
           & NICER                                                 & 0.5-10 keV    & 01:14-01:31  &   6203980121   \\
           &                                                       &               &      &      \\
\textbf{17/09/2023} &                                                       &               &      &      \\
 & HiPERCAM@GTC                                          &     $u_s$,$g_s$,$r_s$,$i_s$,$z_s$   &  20:16-20:24    &  NA    \\
            &                                                       &               &      &      \\

  \hline
&$^*$ Observations started before the midnight, thus on the 08/09.\\
\end{tabular}

\end{center}
\label{tab:log}

\end{table*}

\subsection{HAWK-I Observations} Two IR observations (2023 September 9 and 15) were conducted in the K$_s$-band (2.2$\mu$m) using HAWK-I mounted at the 8.2 UT 4 Yeput telescope at the Very Large Telescope, in Cerro Paranal,  Chile  \citep{pirard}. The near-IR imager is composed of a mosaic of 2$\times$2 Hawaii 2RG 2048$\times$2048 pixels detectors. To achieve high-time resolution, we used the Fast Photometry mode, which reads only one stripe on each detector. In particular, we used a 128$\times$64 window and set the time resolution to 0.125\,ms.  The instrument was rotated to place a reference star (2MASS 17274523-1612197, K$_s$=11.86$\pm$0.03)  in the same stripe.  The observations were divided into data cubes, each made of 250 frames. A gap of a few seconds is present between each data cube. No frame losses were registered during the observations. 

We used an adapted version of the ULTRACAM pipeline software\footnote{\url{https://github.com/trmrsh/cpp-ultracam}} for reading HAWK-I data cubes. We  extract the target count rates using aperture photometry with a seeing-dependent circular aperture tracking the centroid of the source for each frame in the data cube. We normalized the target light curve by the reference star to minimize seeing effects. We found our source to be extremely variable, down to sub-second timescale, with an average root-mean-squared (RMS) of almost 20\%. The average observed fluxes in the two observations changed from 21.5$\pm$0.1\,mJy (September 9) to 22.9$\pm$0.1\,mJy (September 15).

\subsection{ULTRACAM Observations}

High time-resolution  optical imaging was conducted on September 9, 2023 using ULTRACAM on the 3.58\,m New Technology Telescope (NTT) in La Silla, Chile. ULTRACAM was built for fast optical timing in multiple wavebands and includes three channels for simultaneous multi-wavelength monitoring \citep{dhilon2007}.
We observed Swift\,J1727.8--1613 with the custom-made Super-SDSS {$u_s$ (352.6 nm), $g_s$ (473.2 nm), and $i_s$ (771.1nm)} super-SDSS filters. ULTRACAM was used in drift mode with a time resolution of $\approx$16ms. In particular, two sub-windows of the detector, one centered on Swift\,J1727.8--1613 and one on a comparison star, with 54\,$\times$\,54 pixels windows and 2\,$\times$\,2 binning for sensitivity and speed. The target in the $u_s$-band data was found too faint for meaningful analysis and was not considered in this work. 

The data were reduced using the HiPERCAM pipeline software\footnote{\url{https://github.com/HiPERCAM/hipercam}}. The bias was subtracted from each frame and flat field corrections were applied. Aperture sizes scaled to the instantaneous seeing were used. These apertures had variable center positions that tracked the centroids of the sources on each frame, with a two-pass iteration (where an initial pass is made to track the sources on the CCD before a second photometry pass) used for accuracy.
Our comparison star is located at RA = 17:27:49.96, Dec = -16:10:22.65 (J2000), and is listed in The Guide Star Catalog, Version 2.4.2 (GSC2.4.2) \citep{Lasker_GuideStarCatalog242_2008} as S8KB304636 with $g_s$/$i_s$ magnitudes of 13.247$\pm$0.004/12.607$\pm$0.005, respectively. 
The star was taken to be constant and was calibrated using the HiPERCAM images (see Section XX).
We extracted the magnitudes for Swift\,J1727.8--1613  and the comparison star using the same technique described earlier. The aperture also tracked the centroid of the source of interest by using a bright star in the field as a reference. We measured an average flux of 19.00$\pm$0.01 mJy and 11.79$\pm$0.01 mJy in the $g_s$- and $i_s$-bands respectively.  {Although we could not significantly detect the source in the individual $u_s$ band frames, by co-adding all the images we measure a flux of 19.48$\pm$0.01 mJy.} As shown in  \citealt{wild2022}, given the measured $g_s$--$i_s$ colour ($\approx$-0.5), these fluxes have negligible difference from standard sdss filters.

\subsection{HiPERCAM Observations:} High-speed optical imaging was performed on 2023 September 17  using HiPERCAM on the 10.4\,m Gran Telescopio Canarias (GTC) in La Palma, Spain. Similarly to ULTRACAM,  HiPERCAM uses dichroic beamsplitters to simultaneously image the custom-made Super-SDSS {$u_s$ (352.6 nm), $g_s$ (473.2 nm), $r_s$ (619.9 nm), $i_s$ (771.1 nm) and $z_s$ (915.6 nm) filters} \citep{Dhillon21}.
HiPERCAM can observe at frame-rates up to $\sim$1000\,Hz. This is possible due to the lack of a physical shutter and so the frame-transfer CCDs can rapidly shift charge into a storage area for reading out, freeing up the original pixels for observation and thereby achieving low ($\approx $ ms) dead-times \citep{Dhillon21}. The CCDs were binned by a factor of 8 and drift mode was used with two windows of 96$\times$96 pixels each. The instrument was orientated so that one window was centered on Swift\,J1727.8--1613  and another window on a local standard star. We took about 1,300,000 images with an exposure time of 2.87\,ms which resulted in a cadence of 4.129\,ms, allowing us to observe for $\approx$8~minutes starting from 2023-09-17 20:16:32.103 UTC.  Although the presence of a heavy CCD read-out time affects the Fourier analysis close to the Nyquist frequency ($\approx100$Hz) it will not affect the frequency range where the BHTs show most of the power ($\leq$10Hz).

We used the HiPERCAM pipeline software to de-bias, flat-field and extract the target count rates using aperture photometry with a seeing-dependent circular aperture tracking the centroid of the source. We determined the instrumental zero-point using observations of a HiPERCAM standard star \citep{Brown22}, which was then used to  calibrate the local standard star  PSO\,J172749.957-161022.624 (the same star as the ULTRACAM observations), allowing  for the photometric calibration of Swift\,J1727.8--1613. 
The magnitude of the local standard was compared with the values listed in the Pan-STARRS survey DR1 catalogue \citep{Magnier20}  after transformation to SDSS  magnitudes \citep{Finkbeiner16}, and was found to agree at the 10 per cent level.
{The mean observed fluxes of Swift\,J1727.8--1613 are $u_s, g_s, r_s, i_s, z_s$ filters are 12.42$\pm$2.12, 12.89 $\pm$0.48, 14.82$\pm$0.58, 14.73$\pm$0.65, 14.15$\pm$0.81\,mJy, respectively, 
}

\subsection{NICER Observations} We obtained simultaneous X-ray coverage during the first two dates of our O-IR campaign using the Neutron Star Interior Composition Explorer Mission (\textit{NICER}) aboard the International Space Station \citep{nicer2016}. Cleaned events were extracted using the \textsc{heasoft} software \textsc{nicerlc2} using the \textsc{version 11}  and applying the default event-screening conditions. We then extracted the light curves in the 0.5--10\,keV band and binned with a time resolution of 1\,ms using \textsc{xselect}.  

\section{Analysis and results} 

All the optical/near-IR and X-ray light curves show strong stochastic noise across all bands. We first computed the power density spectra (PDS) of each band for each epoch. For the cases where we had strictly simultaneous data for more than one band, we also performed a cross-correlation analysis in both time and frequency domain.

\subsection{Power spectral analysis}

\begin{figure*}
\includegraphics[scale=0.23,angle=-90]{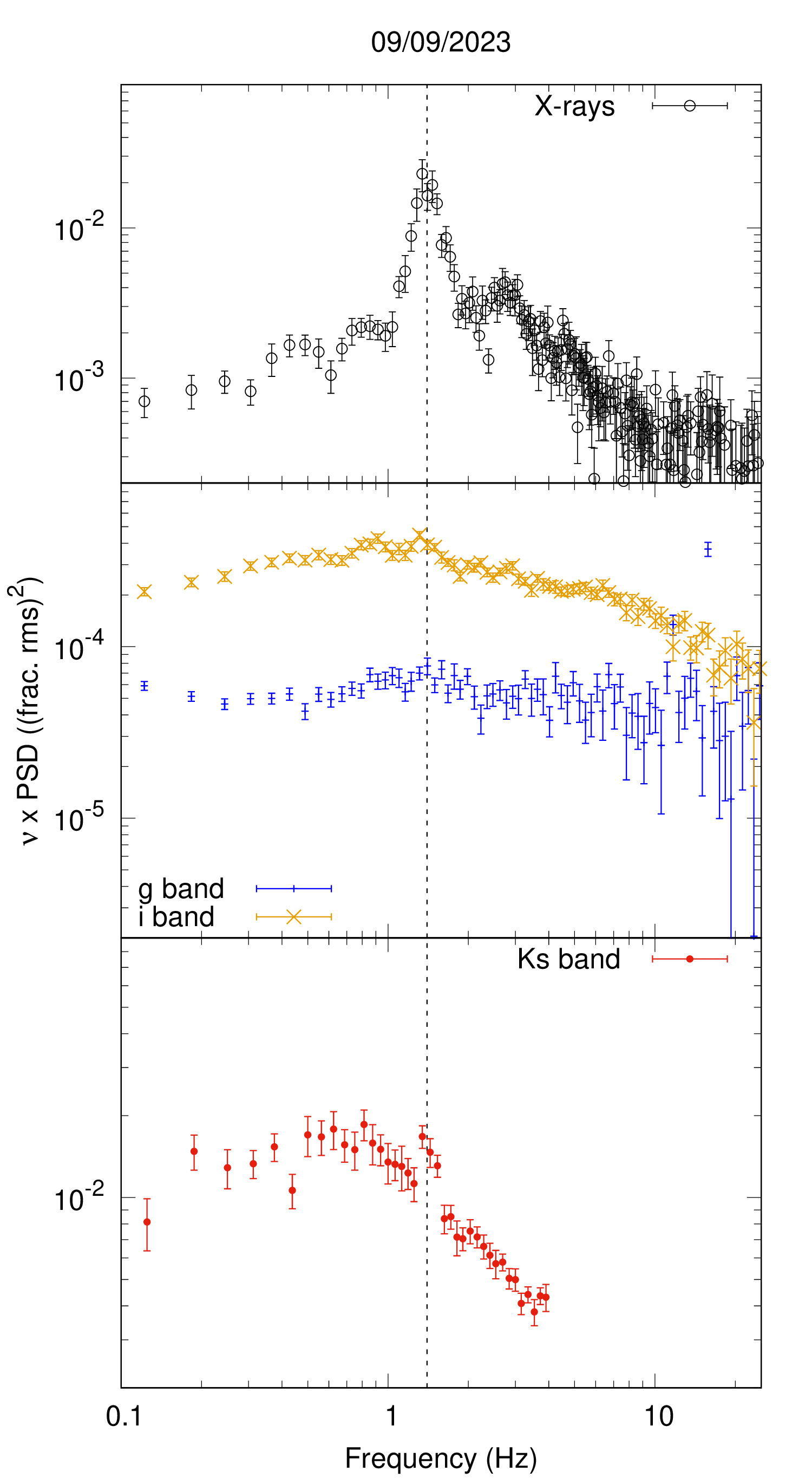}\hspace{-0.25cm}
\includegraphics[scale=0.23,angle=-90]{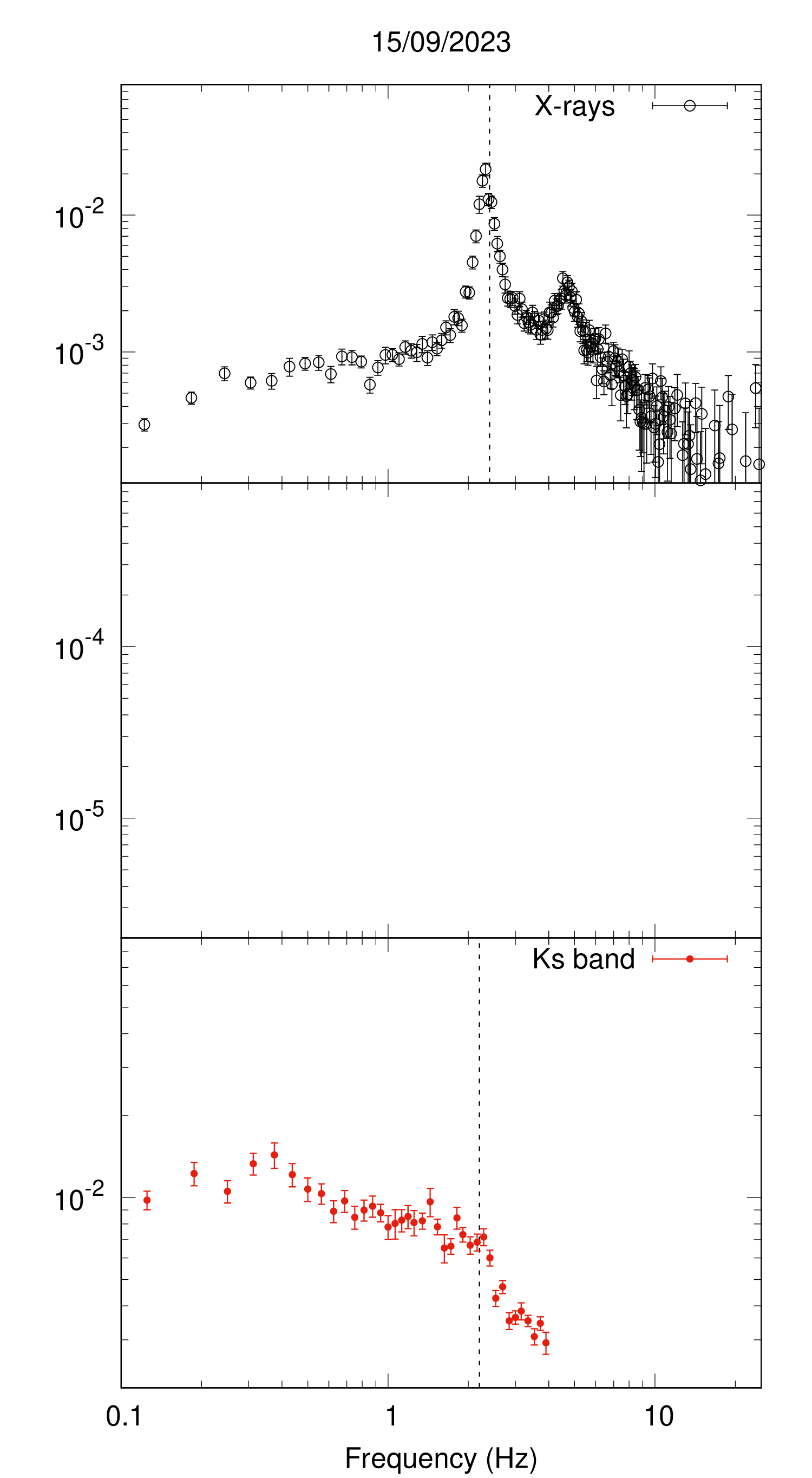}\hspace{-0.25cm}
\includegraphics[scale=0.23,angle=-90]{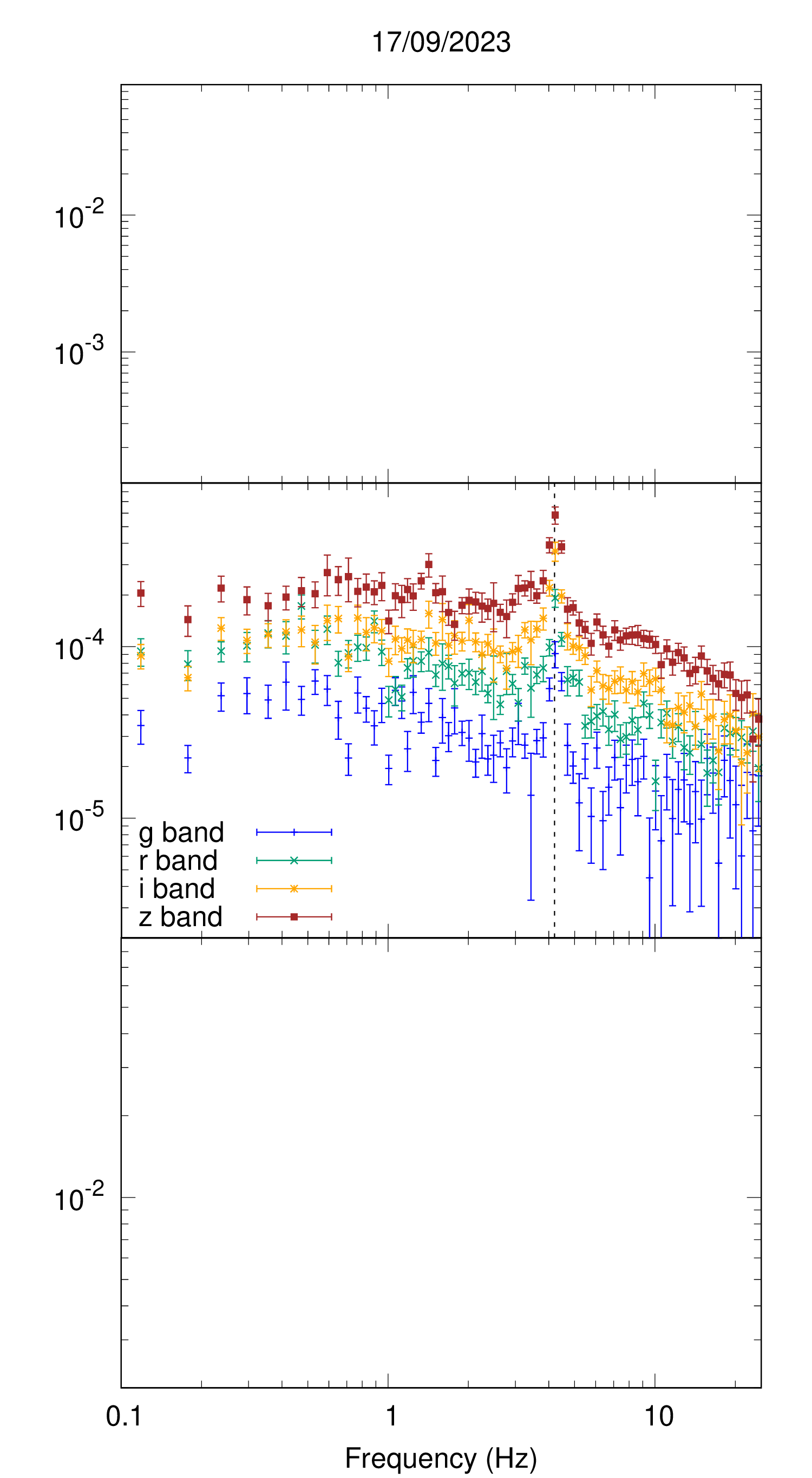}

\caption{Power spectral analysis of our campaign. The \textit{left panel} contains the analysis regarding the first epoch (09/09/2023); the \textit{central panel} the analysis regarding the data of the second epoch (15/09/2023); the \textit{right panel} shows the last power spectral analysis of the last observation with HiPERCAM (17/09/2023). The top panel shows the X-ray PDS, the middle panel shows the optical PDS  and the bottom panel shows the IR PDS. {Vertical dashed lines indicate the frequency of the QPO.}
\label{fig:pds_x-o-ir_evolution}}
\end{figure*}

All the PDS were computed with fractional squared RMS normalization \citep{belloni1990,vaughan2003} and are shown in Fig.\,\ref{fig:pds_x-o-ir_evolution}.  Given that each dataset has a different structure (see Section 2), for this part of the analysis, different numbers of bins per segment as well as logarithmic binning were used.  Here we summarize the different parameters used and the results obtained.
  
\subsubsection{X-rays} We analysed the NICER light curves using 16384 bins per segment and a geometrical rebinning factor of 1.02. The power spectrum shows the typical broadband noise with a strong type-C QPO at 1.4\,Hz, and a harmonic component.  In the second observation, as the source evolved towards a softer state, the QPO increased in frequency to 2.2\,Hz. 
To characterize the evolution of the PDS, we fitted the data using multiple Lorentzian components, as described in \citet{belloni1997b}. Given the low signal-to-noise ratio at high frequencies, we only fitted the data up to 15\,Hz.  The evolution of the properties of the QPO (centroid frequency, RMS, presence of a harmonic component) are reported in Tab.~\ref{tab:fits_pds}. 
Our results are consistent with previous studies of this source using \textit{HXMT} \citep{yu2024,zhu-wang2024} and with the typical evolution of BHTs in the intermediate states.
 
\subsubsection{Optical/near-Infrared}  The ULTRACAM data was analysed using 1024 bins per segment and a geometrical rebinning factor of 1.05. Due to the low statistics at high frequencies, for the fitting procedure, we considered data only up to 15\,Hz. Instrumental peaks visible in $g_s$-band around 10--15\,Hz were also excluded from the data.  Regarding HiPERCAM, we used 4096 bins per segment and rebinning factor of 1.05. We used data beyond 60 Hz to estimate the Poissonian noise level. For the HAWK-I data, we used 128 bins per segment and a logarithmic rebinning of 1.05. As mentioned in the previous sections, the data presented regular gaps every 250 frames. To obtain a better estimate of the lower frequencies, similarly to previous works \citep{kalamkar2016,vincentelli2019,vincentelli2021_1535}, we filled these gaps with a time series with the same count rate, and statistical distribution. In particular, we selected a random segment of the light curve in the previous cube, with the same length of the gap, and inverted it in time. This mathematically keeps the same properties of the original PDS, without introducing any bias.

The O-IR PDS shows a clear QPO at the same frequency observed in the X-rays during the first epoch. The amplitude IR QPO in the second epoch is clearly lower. {During the first epoch (September 09), an optical QPO was detected only in the $i_s$ band. Nonetheless, we obtained an upper limit on the g band QPO, by the forcing the presence of an additional Lorentzian component with the same frequency and width observed in the $i_s$ band. We found a 3 sigma upper limit of 0.25\%.} In optical observations of the last epoch (September 17), we found that the QPO moved up to 4.2\,Hz. This is consistent with the quasi-simultaneous X-ray observations with \textit{HXMT} taken on the same day \citep[][]{yu2024}{}{}. This is the fastest optical QPO ever detected so far for a BH LMXB. As already done for the X-ray PDS, we characterized the O-IR PDS with multiple Lorentzians 
(see Tab.~\ref{tab:fits_pds}).  This analysis confirmed that evolution that the IR QPO decreases with frequency, with a detection only at 3$\sigma$ level in the second epoch.

\subsubsection{The Spectral energy distribution of the O-IR QPO and of the flares} 

\begin{figure*}
\includegraphics[scale=0.7]{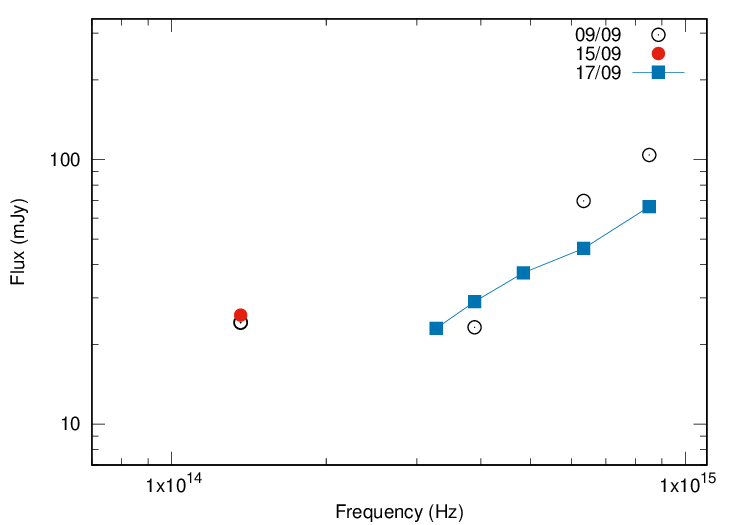} \includegraphics[scale=0.7]{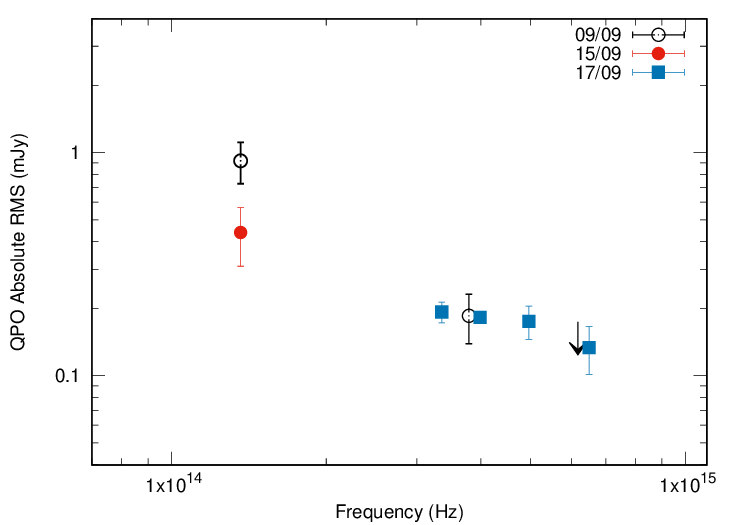}
\caption{ \textit{Left Panel:} Average de-reddened spectral energy distribution of the three epochs. A flat component seems to be present between the IR and the optical band, up to the  $r_s$-band. \textit{Right Panel} Absolute RMS spectrum of the QPO in the three epochs. The RMS increases towards the redder bands, as expected from synchrotron radiation. {From lower to higher energy, the white empty points represent the K$_s$, the $i_s$ and the $g_s$ band; the second epoch is represented only with filled red circles and has only the IR band; for the third epoch with HIPERCAM, the five filled blue squares in the left panel represent (from lower to higher energies)  $z_s$ (915.6 nm), $i_s$ (771.1 nm), $r_s$ (619.9 nm),   $g_s$ (473.2 nm) and $u_s$ (352.6 nm). Given that no timing analysis was possible with the $u_s$  band, the right panel shows only four points. For clarity, the HIPERCAM and ULTRACAM points of the RMS spectrum (right panel) where shifted of few percent on the horizontal axis. }
\label{fig:rms_sed}}
\end{figure*}

Thanks to the simultaneous multi-band observations, we can build the spectral energy distribution (SED) of the O-IR QPO in the different epochs. From the fits reported in 
Table \ref{tab:fits_pds}, 
we extracted the fractional RMS of the O-IR QPO and converted them into absolute units. We derived the line of sight extinction from the \textit{Hubble Space Telescope} high-resolution ultraviolet spectroscopic data obtained on October 2$^{nd}$, applying the method described in \citep{castro-segura_2024}. While a detailed analysis of the data will be presented elsewhere (Castro Segura et al. in prep.), our calculation yields  $E(B-V)= 0.37\pm0.027$ for $R_V = 3.1$.  {The resulting resulting average flux SED and the QPO SED are shown in Fig. \ref{fig:rms_sed} left and right panel respectively}.

The variable SED clearly shows that the overall O-IR variability increases toward the redder bands. The absolute RMS increases from less than 0.2\,mJy in $g_s$- and $r_s$-bands to nearly 1\,mJy in the $K_s$-band (see Fig. \ref{fig:rms_sed}). The QPO amplitude is rather stable at optical wavelengths, while some marginal (less than 2 $\sigma$) variability is observed in the $K_s$-band.

The average SED does has a flat component from the $Ks$-band to the $i_s$-band (or   $z_s$-band in the 3rd epoch) and a steepening at shorter wavelengths. The QPO SED has instead a strong red component from the $Ks$-band to the $i_s$. While no QPO is detected in the $g_s$-band in the 1st epoch, the HiPERCAMs wider spectral coverage and high sensitivity reveal that the QPO SED still has a red slope during the last epoch. 

{
The $\approx$ms resolution of HiPERCAM allowed us to resolve the optical flares in different bands with very high level of detail (Fig.\,\ref{fig:hcam_lcurves}, for subsection of the  observed HiPERCAM light curves). Thus, we took advantage of the unique quality of the dataset to characterize the wavelength-dependent variability also in the time domain. In detail, we identify the flare events by determining the start and end of the same flare event in each waveband. We then subtract the interpolated flux underneath the flare  which in effect subtracts the contribution of the non-variable component, and then sum the flux to give the flux of the flare event in each waveband. We assume during the actual flare event that the other components that contribute to the observed flux do not vary. We clearly identify  11 flare events, defined as events more than 5-$\sigma$ from the local mean value. In order to interpret the broad-band spectral properties of the flares, we compare the observed flare fluxes with synchrotron emission. We first dereddened the wavelength dependent flare fluxes {using a colour excess of E(B-V)=0.37 mag and then fit them with a  power-law form $F_\nu \propto \nu^\alpha$, where $\nu$ is the frequency and $\alpha$ is the spectral index. The obtained flare slope vary from -0.48 +/- 0.24 upto 1.20 +/- 0.26, with an average power-law index of $\alpha$=-0.9$\pm$0.2}.

\begin{figure}
\centering
\includegraphics[scale=0.28]{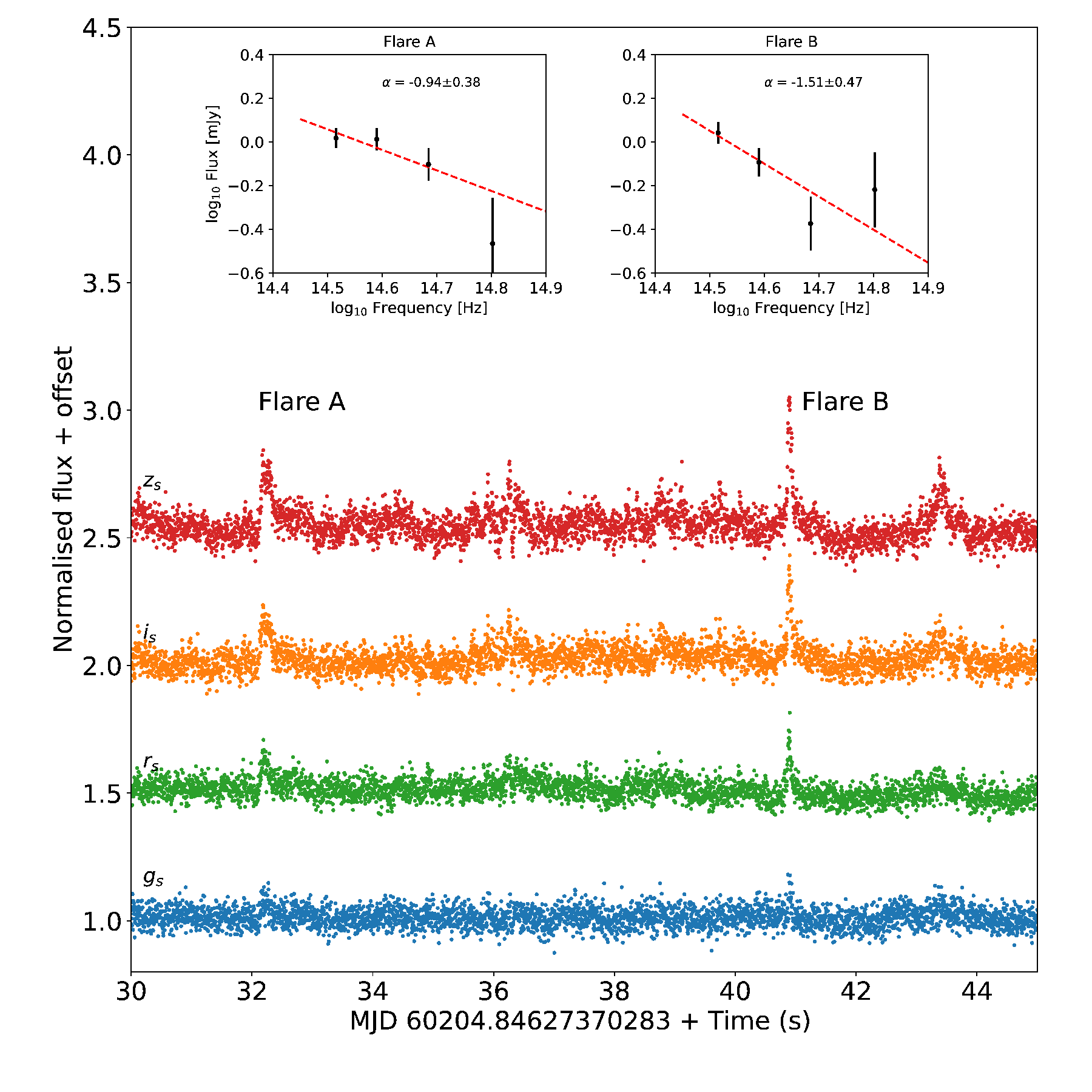}
\caption{A section of the observed HiPERCAM light curves of Swift\,J1727.8--1613. The light curves have been normalised. Strong red flares are clearly seen. The inset show the de-reddened spectral energy distribution of two flare events  using E(B-V)=0.37 mag.
\label{fig:hcam_lcurves}}
\end{figure}

\subsection{Cross-correlation analysis}

\begin{figure*}
\hspace{0.2cm}\includegraphics[scale=0.62]{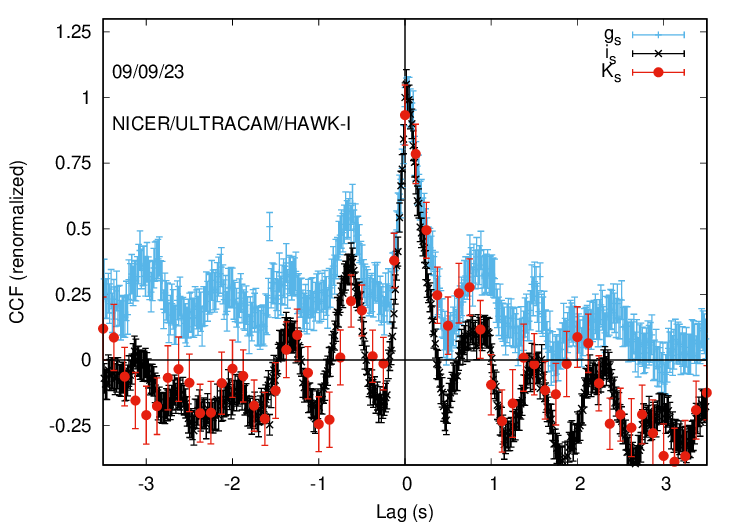} \hspace{-0.2cm}\includegraphics[scale=0.62]{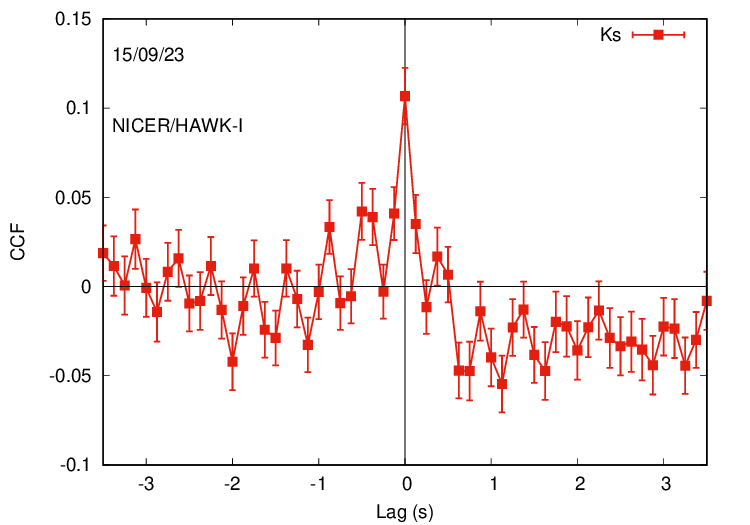}

\includegraphics[scale=0.33,angle=-90]{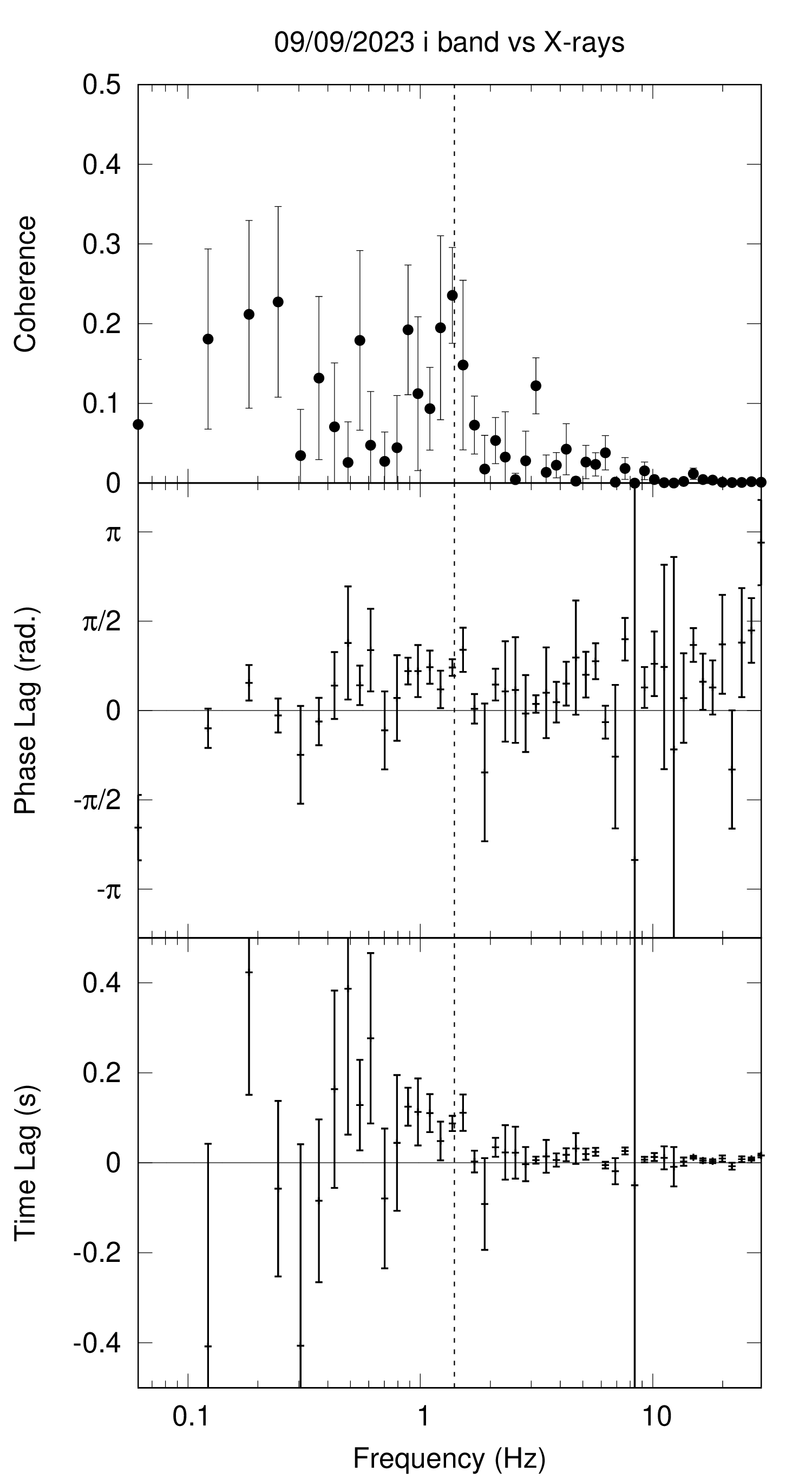}\hspace{-0.5cm}
\includegraphics[scale=0.33,angle=-90]{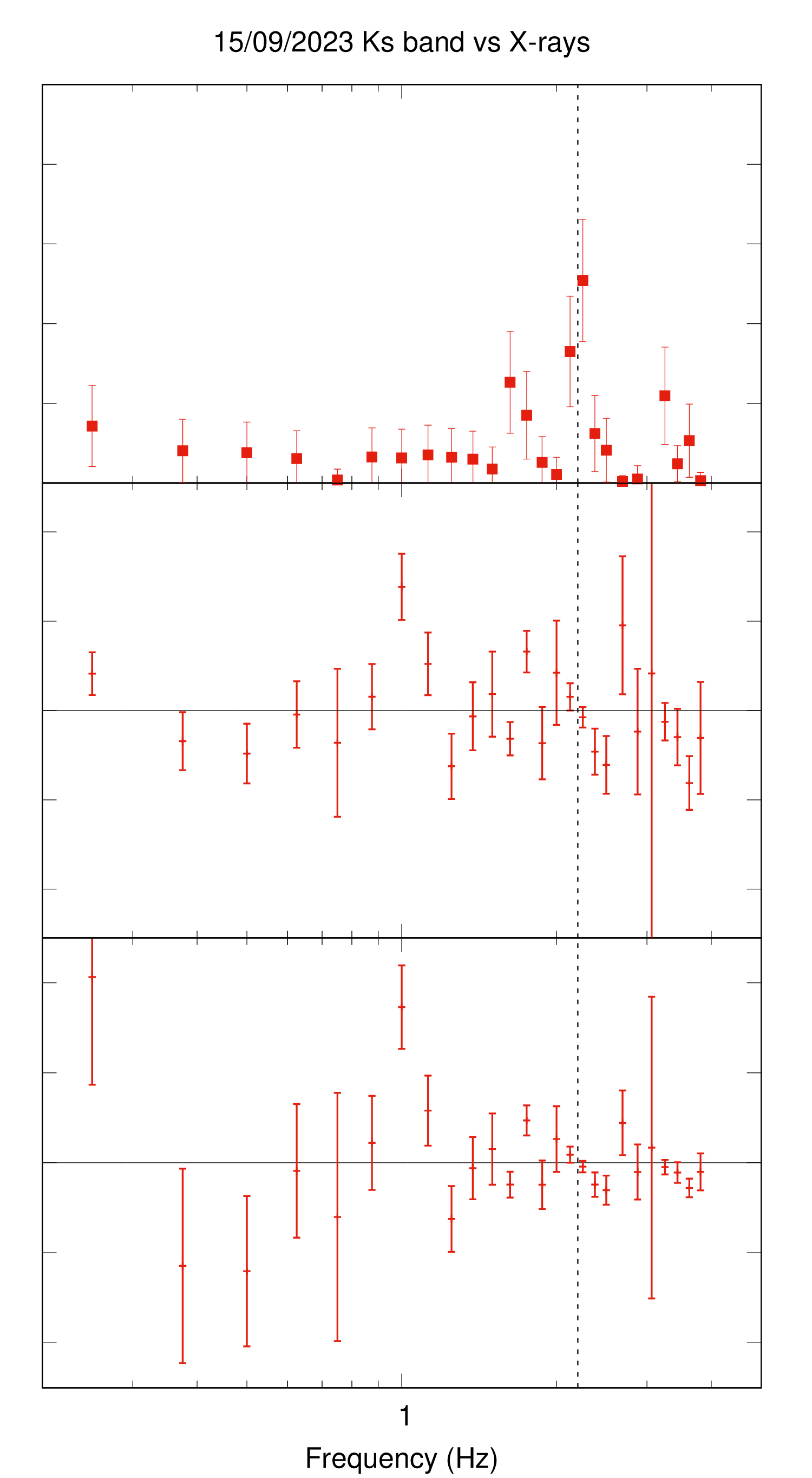}

\caption{The results of the time and Fourier cross-spectral analysis for the data taken during the first two epochs. The left panels contain the analysis regarding the first epoch (September 9), while the right panel the analysis regarding the data of the second (September 15). For all the analysis, we used the X-rays as reference band: thus a positive lag means that the O-IR variability arises \textit{after} the X-rays. The top panels show the cross-correlation function, while the bottom panels show the cross-spectral products (i.e. the results of the Fourier domain analysis). From top to bottom: coherence, phase and time lags. We note that due to the different time resolution of ULTRACAM and HAWK-I, the probed frequency range is different in the left and right panels.  {Similarly to Fig. \ref{fig:pds_x-o-ir_evolution}, vertical dashed line represent the centroid of the QPO from the PDS fitting}.
\label{fig:cross_general}}
\end{figure*}

We performed a cross-correlation analysis on light curves taken during all the epochs. For the cross-correlation function (CCF), we followed the approach described in \citet{gandhi2010}, i.e. we normalized the correlation coefficient using the standard deviation of the data. For the Fourier products, i.e. cross-spectrum, lags and (intrinsic) coherence, we used the steps described in \citet{uttley2014}.

\subsection{09/09/2023} 

This epoch has the largest overlap with all the telescopes involved. We measured 
both ULTRACAM bands and HAWK-I versus the X-rays (i.e. using the X-rays as  reference band), see Fig. \ref{fig:cross_general}, top left panel) and Fig. \ref{fig: zoom}.  All CCFs display a clear, narrow positive peak (i.e. O-IR lagging the X-rays) between 0 and $\approx 100$\,ms. A strong modulation consistent with the QPO frequency can also be seen. We investigated the intraband O-IR lags but found no significant delay (3$\sigma$ limit of 30\,ms). We did not find any significant difference using the different X-ray energy bands.

Given the poorer time resolution of the HAWK-I data, the lower overlap with the X-rays, and the consistency between the optical and infrared bands, we focused the Fourier cross-spectral investigation mainly on the ULTRACAM data. We used the same number of bins per segment and binning factor of the PDS analysis. In Fig. \ref{fig:cross_general} we show the cross-spectral products for only the X-ray versus $i_s$-band, where the overall variability is strongest, and so the signal-to-noise is higher.  An optical lag can be seen at the QPO frequency. An increase in the intrinsic coherence is also seen at the same frequency.  By integrating the cross-spectrum between 1.2 and 1.6\,Hz we measure a lag of 70$\pm$15\,ms (0.62$\pm$11\,rad).  Despite the low coherence and statistics, there is some marginal evidence of a constant lag as a function of frequency below the QPO. To check for additional lag components we also computed the lags integrating the cross-spectrum before and after the QPO. By computing the lags in 0.5--1\,Hz and 2--5\,Hz bands, we find a time lag of 110$\pm$40\,ms, and 15$\pm$6\,ms, corresponding to a phase lag of 0.52$\pm$0.19 and 0.48$\pm$0.18\,rad, respectively.

 \begin{figure}
  \centering
  \includegraphics[width=0.45\textwidth]{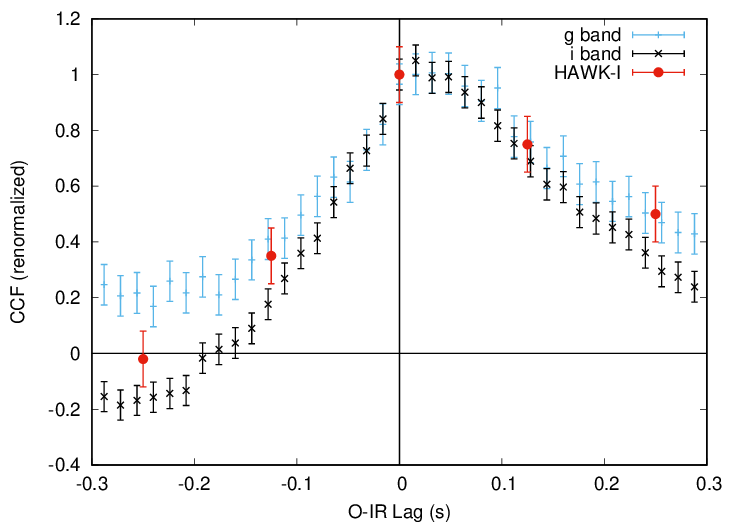}
  \caption{Same as Top left panel in Fig. \ref{fig:cross_general},  but shown between $\pm$0.3\,s. The strong asymmetry demonstrates that a delay between the two bands is present. }        
  \label{fig: zoom}
\end{figure}

\subsection{15/09/2023} 

During this epoch, we only have simultaneous HAWK-I and NICER data. This allowed us to perform analysis both in the time and frequency domain (See Fig. \ref{fig:cross_general}, right panel). However, given the overlap of only $\approx$20 minutes, in order to increase the statistics at higher frequencies, differently from the PDS analysis, we computed the cross-spectrum using 64 bins per segment and a binning factor of 1.05.   We found that the CCF peaks at zero, and no longer shows the $\approx$0.1\,s lag observed on 09/09/2023. The scatter of the CCF is consistent with the effect of a QPO at 2.2\,Hz. The Fourier lags show that at the peak of the QPO, the coherence is maximum, and a zero lag is observed. By integrating the cross-spectrum over the QPO range (2.0 to 2.4\,Hz) we found that the lag is consistent with zero, i.e. -10$\pm$11\,ms (1$-\sigma$). Similarly to the previous epoch we measured the lag over a broad band below the QPO (0.5 to 1.2\,Hz) and found no significant delay (54$\pm$90 ms).
Finally, we investigated again a possible dependence on the X-ray energy, but no significant variations were found.

\begin{figure*}
  \centering
  \includegraphics[scale=0.65,]{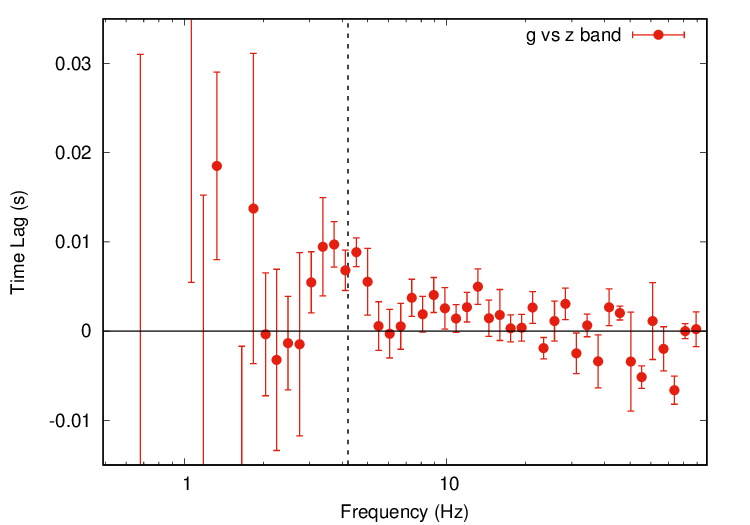}
  \includegraphics[scale=0.65]{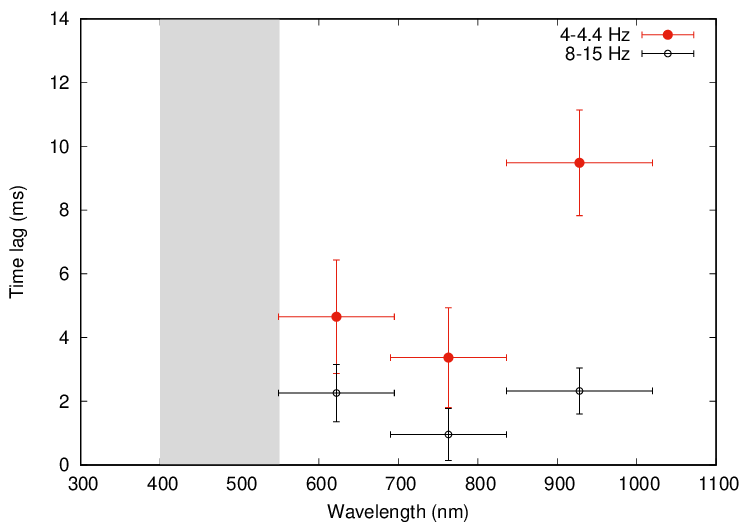}
  \caption{ \textit{Left Panel:}  The lag frequency spectrum between the $z_s$ vs the $g_s$-band computed with HIPERCAM. A positive delay implies that the $z_s$-band lags the $g_s$-band. For guidance, we show in grey the rescaled the $z_s$-band PDS. The observed lag is clearly linked to the QPO. \textit{RIght Panel:} Lag vs wavelength spectrum for the QPO (red points) and the broadband noise(black points). The grey area shows the $g_s$-band, which was used as reference band. }
  \label{fig:hipercam}
\end{figure*}

\subsection{17/09/2023} 

No X-ray coverage was achieved for this epoch. Thus, the cross-correlation analysis can be performed only using the different optical bands in the HiPERCAM data. As a first test, we investigated the presence of lags between the $g_s$- and $z_s$-bands from the  Fourier cross-spectrum, as done for the first two dates.  We used the same number of bins per segment and binning factor of the PDS analysis. As shown in the left panel Fig.\,\ref{fig:hipercam}, a clear excess is present at the QPO frequency, with a delay of the $z_s$-band of roughly 10\,ms.  The fact that the signal appears only around the QPO frequency excludes that the delay is due to an instrumental unknown systematic  We then proceeded to compute the lag-wavelength spectrum at the QPO frequency: i.e. evaluating the phase lag after integrating the cross-spectrum between 4 and 4.4\,Hz, and compared it with a lag at higher frequencies (Fig. \ref{fig:hipercam}, right panel). We found two main results: a smaller lag is also present using the $r_s$- and $i_s$-band, suggesting that the delay increases with wavelength.  We also found a constant lag at high frequency of $\approx$ 2\,ms. Given that this lag is close to the time resolution, this may be a systematic effect due to the dead time.
 
\section{Discussion} \label{sec:cite}

We monitored the rapid multi-wavelength variability of the black hole transient Swift\,J1727.8--1613 with three observations during the hard-to-soft transition. The source was highly variable in all bands: through Fourier analysis, we discovered the presence of an O-IR QPO following the frequency of the X-ray one. The Fourier and time domain analysis suggest that the correlation seen in the CCF arises mainly from the QPO. In the first epoch, there is some marginal evidence for lags at frequencies below the QPO, but there is insufficient signal-to-noise to quantify the contribution of the broadband noise. More importantly, the lag at the QPO frequencies changes between the first two epochs, going from $\approx$70 ms to 0. Finally, by analysing the HiPERCAM data we measured for the first time an intraband lag only at the QPO frequencies, with $g_s$ lagging $z_s$ by $\approx$10\,ms.

 In X-ray binaries, three components are expected to have a significant contribution in the O-IR band: thermal reprocessing from the outer disc or donor star and synchrotron radiation from either an extended hot flow or the jet \citep{veledina2011,malzac2014}. A superimposition of these components is often present in the O-IR regime, and with only a few photometric bands it is impossible to disentangle them. However, these components exhibit different spectral shapes: thus the analysis of the SED variability can help us to pinpoint the origin of the O-IR QPO.

\subsection{Disc reprocessing}

Thermal reprocessing from the outer disc could generate both an O-IR QPO and an O-IR flux excess. Thus in principle, it may be possible to produce the observed flat SED (\citealt{hynes2005,zurita-heras2011,gierlinski2009}). However, a few  elements allow us to discard this scenario. First, the variable thermal emission from an irradiated disc is expected to exhibit a blue or flat spectrum rather than the red slope observed in our data \citep{gierlinski2009}. Furthermore, thermal reprocessing from the outer part of the disc should also produce a lag of the orders of seconds, which we do not observe. Finally, the outer parts of the disc or the donor star are too large to produce sub-second variability via reprocessing: the emitting region is expected to be of the order of a few light seconds, thus the width of such a response function would smooth away any variability faster than $\approx$1\,Hz \citep{obrien,veledina2015}. 

We conclude therefore that most of the observed variable O-IR emission must be non-thermal. In particular, comparing also to previous O-IR BH LMXBs flaring studies, the $\alpha\approx-0.9$ measured with HiPERCAM's fast flares seems to be consistent with cyclo-synchrotron radiation  \citep[see e.g.][and references therein]{gandhi2016,Shahbaz2023}. Thus, in the next section, we will discuss the implications of our results regarding synchrotron radiation from the hot flow and the jet.

\subsection{Hot flow}

A geometrically thick, optically thin hot magnetized flow can emit in the O-IR regime through synchrotron radiation from its outer regions \citep{veledina2013}.  Similarly to jets, the superposition of partially absorbed spectra from different regions with a decaying magnetic field will give rise to an O-IR excess. The spectrum will then decay steeply below the characteristic frequency associated with the truncation radius of the hot flow.  According to our SED, the break should then lie  around $\approx 10^{14}$ Hz (i.e. in the IR band).   As shown by previous studies \citep{vincentelli2021_1535}, this component can successfully reproduce a $\approx$1--2\,Hz QPO with a truncation radius of a few tens of gravitational radii. 

If the disc truncation radius is large enough, synchrotron radiation can be the main source of seed photons for the Comptonized emission \citep{wardzinskizdziarski2001,poutanen2009}. The same population of electrons would be responsible for Comptonization and synchrotron radiation. This would lead to an anti-correlation between the X-ray and optical-infrared broadband noise \citep{veledina2011}, which is not observed in the data.

Interestingly, depending on the parameters, the hot flow scenario could also explain a correlation between X-ray and O-IR emission. As discussed in \citet{ulgiati2024} a positive correlation is possible if the observed O-IR band falls at frequencies lower than the synchrotron break ($\nu_{O-IR}<\nu_{break}$). Our data shows that the break is beyond the IR band ($\nu_{IR}$> $\nu_{break}$), thus, this particular scenario seems to be excluded.  Along with this, a correlated signal may also arise if the source of seed photons is not synchrotron, but thermal emission from the disc \citep{veledina2018}. This means that if the broad variability plays a significant role, synchrotron radiation must be produced after the thermal Comptonization process. 

Regarding the QPO mechanism, instead, calculations assuming a hot flow undergoing Lense-Thirring precession \citep{veledina_poutanen_ingram_2013} showed that the different emission profiles of the optical and X-ray emitting regions lead to an anti-correlation of their light-curves for low system inclinations (i.e. phase lag of $\pm\pi$) and to a correlation with zero lag at high inclinations. While the high-inclination case could explain the QPO lag observed in the second epoch, neither prediction matches the QPO lag from the first epoch. This indicates that an additional component is required if the hot flow is contributing to the emission. Further modelling for QPO above 1 Hz, beyond the aim of this paper, is required to adequately constrain this component

\subsection{Relativistic jet}

The emission from a jet is expected to dominate the SED towards the redder bands and can naturally explain the observed flux and RMS spectrum. At first glance, the shape of the CCF of the first epoch seems to be consistent with the typical 0.1-second lag usually associated with the jet   \citep[see e.g.][]{gandhi2008,gandhi2017,casella2010,malzac2014}. However, the frequency-resolved analysis reveals a behaviour that is not straightforward to interpret. Time lags and coherence show that most of the signal observed in the CCF arises from the QPO. Past observations, instead, have shown that the lag associated with the jet is observed over a broader range of frequencies,  between $\approx$0.1 and 10\,Hz \citep{gandhi2010,vincentelli2018,vincentelli2019,paice2019,ulgiati2024}.
Thus, it is not clear if we can directly associate  the 70\,ms lag we measure over the QPO frequency range with the 100-ms lag observed over a broader frequency range in other sources.

If the O-IR emission arises from the jet, a possible way to generate a QPO is through precession \citep{kalamkar2016,malzac2018,liska2018}.  The lag between the X-ray and the OIR emitting region should be related somehow to the propagation time of the inflow perturbation to reach the O-IR jet emitting region. However, this would imply that the same lag should be observed in the broad-band stochastic noise. We only have marginal evidence for a lag of $\approx$100\,ms between 0.5 and 1\,Hz, and for a lag of $\approx$15\,ms in the 2--5\,Hz range. Given the data quality, we cannot determine if these features can be associated with the \lq\lq classical\rq\rq~ jet lag over a broad range of frequencies. However, if the 70ms delay is arising only with the QPO and not the broadband noise, then there must be a mechanism to cancel out the \lq\lq classical\rq\rq~ 0.1\,s jet lag. 

To date, it is still unclear how internal shocks regulate the appearance of the correlation and amplitude of the delay of such a delay. Simulations have shown that the amplitude of the jet lag is mainly affected by the inclination angle between its axis and the observer, while the jet Lorentz factor ($\Gamma$) mainly affects its coherence \citep{malzac2018}. From an analytical point of view, the lag is also function of the  power spectrum of the driving fluctuations \citep[Vincentelli \& Malzac in prep; ][]{malzac2014}. A combination of low inclination angle, high $\Gamma$  and high power spectral break frequency could therefore potentially explain why the lag is not well detected. In this context - should this prediction be confirmed by simulations - our measured O-IR time lag would be informative on the structure of the jet.

We also noticed that the changes in lag and RMS are hard to interpret. Lag at the QPO should be linked to the projected distance between the X-ray and the IR-emitting region \citep{veledina_poutanen_ingram_2013,malzac2018}.  However, a variation in the observed lag could also be due to a change in the properties (lag/rms) of the X-ray QPO;  recent studies on this source have shown that the phase lag spectrum changes of less than 10\,ms in this frequency regime, and thus cannot take into account alone the observed variation \citep{zhu-wang2024}. {Moreover, HMXT observations have also revealed the presence of a QPO up to 200 keV. The properties of QPOs at such high energies are not fully understood \citep{huang2018,ma_nat,ma2023}. While they might be explained with complex angular distribution of the hot flow, a quantative model is still missing. Nonetheless, the main spectral-timing features such QPOs has been explained in terms of a precessing jet-like corona \citep{ma_nat}. A variation in the X-ray/IR lag would suggest that the distance between the X-ray and IR emitting region is changing (i.e. the inner structure of the jet is changing).} 
Interestingly, past broadband spectral studies of BH LMXBs show that the break associated to the jet is also shifting in frequency during the transition \citep[see e.g.][]{russell_t2013,russell2020,Echibur2024}. Yet radio/sub-mm observations of this Swift J1727 seemed not to show any strong evolution during our campaign, \citep[][Hughes et al. in prep]{ingram2024}. Further observations, combining X-ray/IR lag measurements and the broadband SED observations with a more frequent sampling, will be necessary to fully understand the evolution of these systems.

Regarding the RMS spectrum, if we assume that the modulation is due to Doppler boosting, the amplitude of the QPO should scale with the jet speed \citep{malzac2018}. The observed variation of a factor of 2 in the RMS requires a very strong change in velocity between the two epochs. Given that we do not observe a dramatic change in the radio or O-IR luminosity, such a scenario would be unlikely. A more probable solution may be the presence of an additional mechanism which decreases the coherence of the QPO. Further monitoring of these sources at high Fourier frequencies is required to characterize this behaviour fully.

\subsection{On HIPERCAM's intraband optical lag}

One of the most peculiar features that emerged from our campaign is the discovery of an optical intra-band lag at the QPO frequency. This is not the first time that such a lag between different optical bands is observed in a BHT. A similar lag between $g_s$- and the $z_s$-band of $\sim$15\,ms in the BHT MAXI\,J1820+070 \citep{paice2019}.  However, this delay was associated with the broadband noise, while in our case the lag is only observed at the QPO.   \citet{paice2019}  interpreted the delay in MAXI J1820+070 as a signature of a stratification of the electrons along the jet stream: such behaviour is expected in the \citet{blandford&konigl} scenario, according to which, the flat radio spectrum of jets is due to the superposition of synchrotron radiation from different population of electrons at different energies along the stream.

 The red slope from SED of the whole variability and of the QPO (Fig. \ref{fig:rms_sed} and Fig. \ref{fig:hcam_lcurves}), clearly showed is non-thermal. This suggests, as in the previous nights, that the emission comes from the hot flow or the jet.  In particular, for the case of the jet, in order produce the emission from the different O-IR bands non-simultaneously, some kind of stratification is required. If a stratified jet is precessing as a solid body, then a light travel time distance delay between the emitting regions is expected. However, as mentioned above, in the case of solid body precession, a lag due to a separation between the $g_s$ and the $z_s$-band emitting region would be constant as a function of Fourier frequency.  This suggests that other processes are at work; for example a non-solid body precession (e.g. if different parts of the jet precesses with different orientations). In that case, a phase lag at the QPO frequency (on top of the propagation timescale) would be directly linked to the  angle of the jet's kink. More in general, this means that the process driving the lags for the QPO is different than the one from the broadband noise. It is clear that a more detailed model to reproduce multi-wavelength lags in jets is required to make more quantitative constraints.
 
Regarding the jet origin, a delay between the different optical bands has been recently envisioned within the jet instability scenario proposed by \citet{ferreira2022}. According to the authors, if some kind of instability triggers jets near a magnetic recollimation zone, such a perturbation (or \lq \lq wobbling\rq\rq)  would travel both up and downstream. New simulations are required to confirm the presence of such a lag and to constrain its amplitude. Thus, although this process has still to be confirmed by simulation, similar measurements of  O-IR time lag would be able to probe the structure of the jet.

Finally, we also note that X-ray polarimetric observations seem to indicate that the source has, most likely, a horizontally extended corona \citep{Veledina2023,ingram2024,Podgorny2024}. As discussed above, it has been shown that if the truncation radius of the flow is large enough, this component can also emit at O-IR wavelengths through synchrotron radiation, especially in the intermediate states \citep[][]{wardzinskizdziarski2001,poutanen2009,veledina2013,veledina2017}{}{}.  If the O-IR QPO signal arises from the accretion flow, a standard solid body precession of the whole flow should not give rise to an O-IR intraband delay. A more complex geometry, which includes some kind of warping, also foreseen by GRMHD simulations \citep[][]{liska203}{}{}, may give rise to an O-IR wavelength dependent lag. More detailed models of accretion discs are necessary to confirm this hypothesis.

\section{Conclusions}

We studied the evolution of the X-ray and O-IR QPOs in the newly discovered black hole transient Swift\,J1727.8--1613 with three multi-wavelength timing observations taken during the hard-intermediate X-ray state. Our results can be summarized as follows:

\begin{itemize}

\item 
The X-ray and O-IR type-C QPOs evolved in frequency in a consistent manner, increasing from 1.4 to 4.2 Hz. The RMS amplitude of the O-IR QPO was stronger in the redder bands, suggesting a non-thermal emission. While the QPO RMS spectrum seems to increase starting from the $r_s$-$i_s$ band across our campaign, the average SED does not show a relatively flat component from r to the Ks band.

\item The O-IR vs X-ray lag at the QPO frequency shifted from $\approx$70\,ms at 1.4\,Hz to zero at 2.2\,Hz. However, current state-of-the-art models which invoke hot magnetized flow or jet cannot easily reproduce this behavior. This may be due to fact that the aforementioned models have been initially developed to reproduce QPOs with lower frequencies, and may, thus, be lacking of some additional physical ingredient.  Further exploration of the parameter space of these models are required, to adequately understand this phenomenon. 

\item 
When the QPO reached 4.2\,Hz we measured an optical intraband lag of $\approx$10\,ms between the $g_s$ and the $z_s$-band, consistent with previous optical intraband lags observed in of MAXI\,J1820+070 \citep{paice2019}.  Yet, this is the first time an intraband lag is measured only at a QPO frequency. Regardless of the physical component producing the oscillation, if it is due to precession, the data suggest that the emission is somewhat anisotropic.
\end{itemize}

Our results demonstrate that during the hard-to-soft transition, both the SED and the O-IR properties evolve over a few days. Future multi-wavelength campaigns with higher cadence and multi-epoch monitoring are required to fully unveil the evolving phenomenology of BHTs. 

\section*{Acknowledgments}

{We thank the referee for the constructive comments which improved the quality of this paper.}
HiPERCAM was funded by the European Research Council under the European Union’s Seventh Framework Programme (FP/2007-2013) under ERC-2013-ADG Grant Agreement no. 340040 (HiPERCAM), with 
additional funding for operations and enhancements provided by the UK Science and Technology Facilities Council (STFC).
FMV and TS acknowledge support from the Agencia Estatal de Investigación del Ministerio de Ciencia, Innovación
y Universidades (AEI-MCIU) under grant PID2020-114822GB-I00 and PID2023-151588NB-I00. FMV and DA acknowledge support from the Science and Technology Facilities Council grant ST/V001000/1. FMV acknowledges the financial support from the Visitor and Mobility program of the Finnish Centre for Astronomy with ESO (FINCA). VSD is supported by STFC.
AJT acknowledges the support of the Natural Sciences and Engineering Research Council of Canada (NSERC; funding reference number RGPIN-2024-04458). PG acknowledges support from a Royal Society Leverhulme Trust Senior Fellowship and support from an STFC Small Award
This research has been supported by the Academy of Finland grant 355672 (AV). Nordita is supported in part by NordForsk.

\section*{Data Availability}

Raw NICER event files are available on the \textsc{heasarc} archive. HAWK-I raw data are available on the ESO public archive. ULTRACAM and HiPERCAM can be made available
upon reasonable request to the authors.  


\bibliographystyle{mnras}
\bibliography{example} 

\appendix

\section{Power density spectra fit}

Tab. \ref{tab:fits_pds} reports the results of the PSD fit using Lorentzian components for each epoch and band.  {We also show the resulting fit and residuals for each PDS in Fig. \ref{fig:nicer_res}, \ref{fig:hawki_res}, \ref{fig:ucm_res}, and \ref{fig:hpcm_res}.}

   \begin{table*}
   
    \caption{Parameters of the fit to the PSD with multiple Lorentzian components, $L(f)=A [\Delta/ (2\pi -atan(2f_0/\Delta))]/ [(f-f_0)^2+(\Delta/2)^2]$, as defined in \textsc{xspec}. We additionally define Q=$f_0/\Delta$.
    The fractional rms of each Lorentzian was obtained from the squared of the integration over the whole frequency range .}
    \label{tab:fits_pds}
    \begin{tabular}{ccccccccc}
     \hline

Date & Band &  Comp. & $f_0$ & $\Delta$ & A  & rms (\%) & Q & $\chi^2/$d.o.f.\\
  & &    & (Hz) & (Hz) &  ($10^{-3}$) &  & & \\
     \hline
   \newline

 09/09/2023 & X-rays & 1&  0 (fixed) & 0.94 $\pm$ 0.15  & 3.8 $\pm$0.3 & 4.3   $\pm$0.4  & -- & 128/121\\
  &         &  2 &  4.1 $\pm$ 0.4 & 3.3 $\pm$ 0.5 & 0.9 $\pm$ 0.3 & 2.8 $\pm$  0.8 & 1.3 $\pm$ 0.2& \\

       &         &   3 & 1.39 $\pm$ 0.01 & 0.29 $\pm$ 0.03 & 6.1 $\pm$ 0.4 & 7.6  $\pm$  0.2 & 4.8 $\pm$ 0.5& \\
       
       &         &   4 & 2.84 $\pm$ 0.04 & 0.67 $\pm$ 0.12 & 1.0 $\pm$ 0.1 & 2.7  $\pm$  0.1 & 4.0 $\pm$ 0.6& \\
\newline
&&&&\\
    & $g_s$ band  & 1&  0 (fixed) & 0.07 $\pm$ 0.01  & 0.29 $\pm$0.05 & 1.2 $\pm$0.2  & -- & 57/56\\
    &  & 2&  0 (fixed) & 2.6 $\pm$ 0.3  & 0.17 $\pm$0.01 &0.9 $\pm$0.1  & -- & \\
 \newline
&&&&\\
     & $i_s$ band  & 1&  0 (fixed) & 0.7 $\pm$ 0.1  & 0.86 $\pm$0.03 & 2.1   $\pm$0.2  & -- & 66/58\\
    &  & 2&  0 (fixed) & 7.6 $\pm$ 0.4  & 0.60 $\pm$0.02 &1.7  $\pm$0.2 & -- & \\
    &  & 3&  1.33$\pm$0.02 & 0.4 $\pm$ 0.1  & 0.8$\pm$0.2 & 0.8 $\pm$0.2  & 3.3   $\pm$ 0.8  \\
     &  & 4&  0.86$\pm$0.02 & 0.24 $\pm$ 0.08  & 0.06 $\pm$0.02 &0.8 $\pm$0.3  & 3.6   $\pm$ 1.1& \\
 
 \newline
&&&&\\
     & Ks band  & 1&  0 (fixed) & 0.94 $\pm$ 0.03  & 4.9 $\pm$0.1 & 15.7   $\pm$0.1  & -- & 29/25\\
    &  & 2&  1.39$\pm$0.02 & 0.15 $\pm$ 0.06  & 1.5 $\pm$0.3 & 3.8   $\pm$0.8  & 9.3$\pm$3.7 & \\

   \\   
    \hline
   \newline

     15/09/2023 & X-ray &   1 & 0 (fixed)      &    1.1 $\pm$0.1  & 2.1  $\pm$  0.2 &  3.2  $\pm$  0.3& - & 131/121\\
       &  &   2 & 2.7$_\pm$0.4    &    4.7 $\pm$0.4     & 1.8 $\pm$  0.2 & 3.8 $\pm$  0.1 & 0.7$\pm$0.2 &  \\
       &  &   3 & 2.31$\pm$0.01    &    0.25 $\pm$0.03  & 3.1 $\pm$  0.3 &  5.7$\pm$0.1 & 9.3$\pm$0.4 \\
       &  &   4 & 4.59$\pm$0.01    &    0.73 $\pm$0.1   & 0.45 $\pm$  0.6 & 2.1$\pm$0.3& 6.3$\pm$0.9     \\
 
 \newline
&&&&\\

       &Ks band &   1 & 0 (fixed)      &   0.49 $\pm$0.09  & 39.1  $\pm$  0.2 &  13.8  $\pm$  0.2 & - & 22/24\\
       &  &   2 & 1.05 $\pm$0.20   &    2.03 $\pm$0.15   & 6.9 $\pm$  0.2 &  7.2  $\pm$  0.1  &  0.5 $\pm$  0.1  &  \\
       &  &   3 & 2.22$\pm$0.03    &    0.14 $\pm$0.09 & 0.3 $\pm$  0.1& 1.7$\pm$0.5 & 16 $\pm$  10 &   \\

   \\   
    \hline
   \newline

     17/09/2023 & $g_s$ band &   1 & 0 (fixed)      &    6.9 $_{-1.8}^{+6.7}$  &5.3$\pm$  0.2 ($\times$10$^{-2}$  ) & 0.51$\pm$0.02 & - & 56/48\\
       &  &   2 & 0.22$_{-0.21}^{+0.07}$    &    0.75 $_{-0.19}^{+0.27}$ & 0.11 $\pm$  0.01 & 0.9$\pm$0.1 & - &  \\
       &  &   3 & 4.24$\pm$0.02    &    0.18 $\pm$0.14 & 0.9$\pm$  0.2 ($\times$10$^{-2}$  ) & 0.29$\pm$0.07  & 24$\pm$18 &  \\

 \newline
&&&&\\

  & $r_s$ band &   1 & 0 (fixed)      &   6.2 $_{-1.8}^{+3.3}$  &8.2$\pm$2.0 ($\times$10$^{-2}$  ) & 1.6$\pm$0.7  & - & 52/46\\
       &  &   2 & 0 (fixed)  &    0.7 $\pm{0.1}$ & 0.34 $\pm$  0.04 &   1.2 $\pm$0.2 & - &  \\
       &  &   4 & 4.22$\pm$0.03    &    0.46 $\pm$0.12 & 2.0 $\pm$  0.4 ($\times$10$^{-2}$  ) & 0.47$\pm$0.08 & 9.2$\pm$2.4 &  \\
       
 \newline
&&&&\\

  & $i_s$ band &      1 & 0 (fixed)      &   8.3 $_{-3.7}^{+4.5}$  &0.13$\pm$  0.03 & 0.8$\pm$0.1 & - & 53/46\\
       &  &   2 & 0 (fixed)   &    1.49$\pm$0.35& 0.34 $\pm$  0.03 & 1.3$\pm$0.1  & - &  \\
       &  &   3 & 0 (fixed)     &    0.12 $\pm$0.07 & 0.17$\pm$  0.03 & 0.9$\pm$0.2& - &  \\
       &  &   4 & 4.18$\pm$0.03    &    0.54 $\pm$0.12 & 4.0 $\pm$  0.1 ($\times$10$^{-2}$  ) &  0.63$\pm$0.02 & 7.7$\pm$1.4 &  \\

 \newline
&&&&\\

  & $z_s$ band &   1 & 0 (fixed)      &    0.20 $\pm$0.05  &0.46 $\pm$  0.01   & 1.5$\pm$0.1 & - & 40/44\\
       &  &   2 & 0 (fixed)    &    2.0$_{-0.5}^{+0.2}$ & 0.54 $\pm$  0.06 &1.6$\pm$0.2& - &  \\
       &  &   3 & 0 (fixed)  &    15.0 $\pm$2.5 & 0.20$\pm$  0.03   & 1.0$\pm$0.1 & - &  \\
       &  &  4 & 4.19$\pm$0.02  &    0.41 $\pm$0.08 & 7.3$\pm$  0.8 ($\times$10$^{-2}$  )    & 0.84$\pm$0.09 & 10.2$\pm$1.9 &  \\

 \newline

    \end{tabular}
   \end{table*}


\begin{figure*}
  \centering
  \includegraphics[scale=0.5]{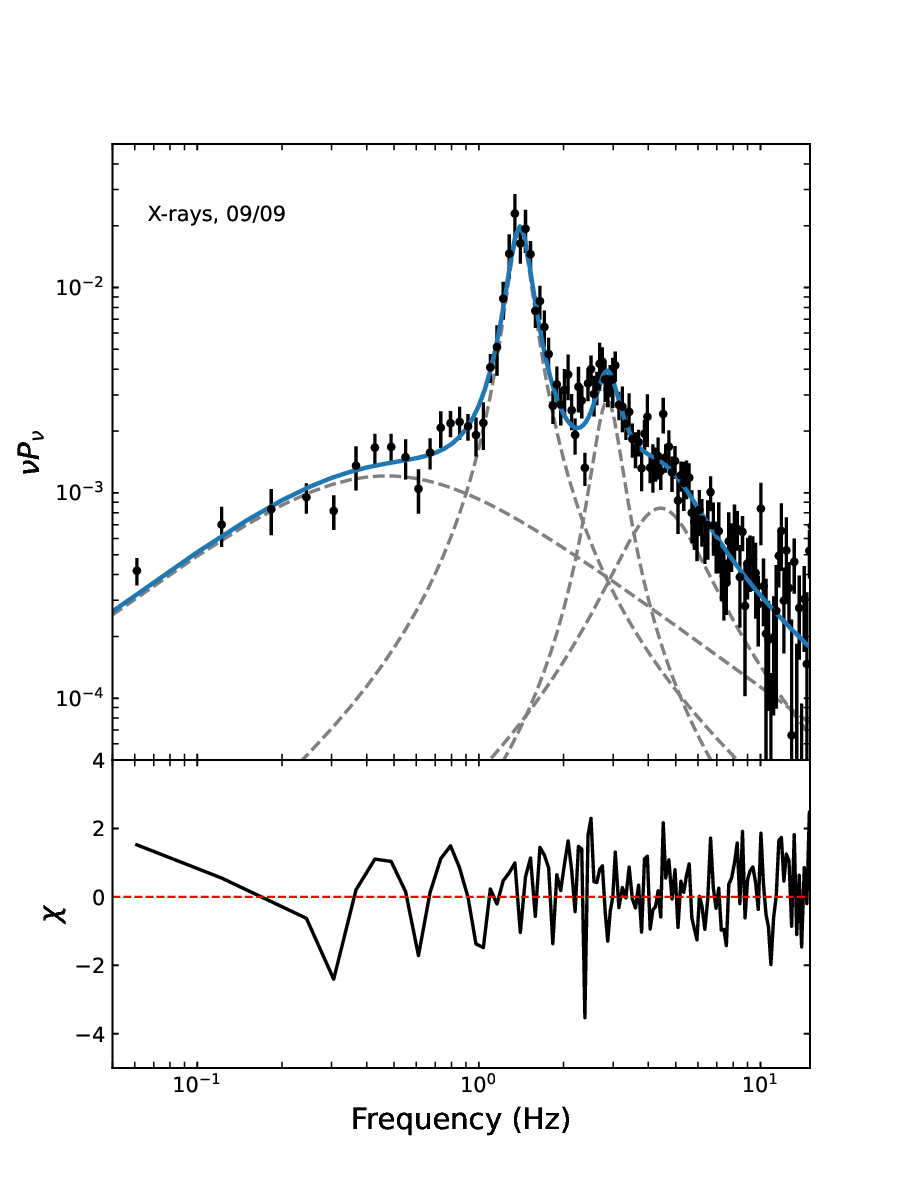}
  \includegraphics[scale=0.5]{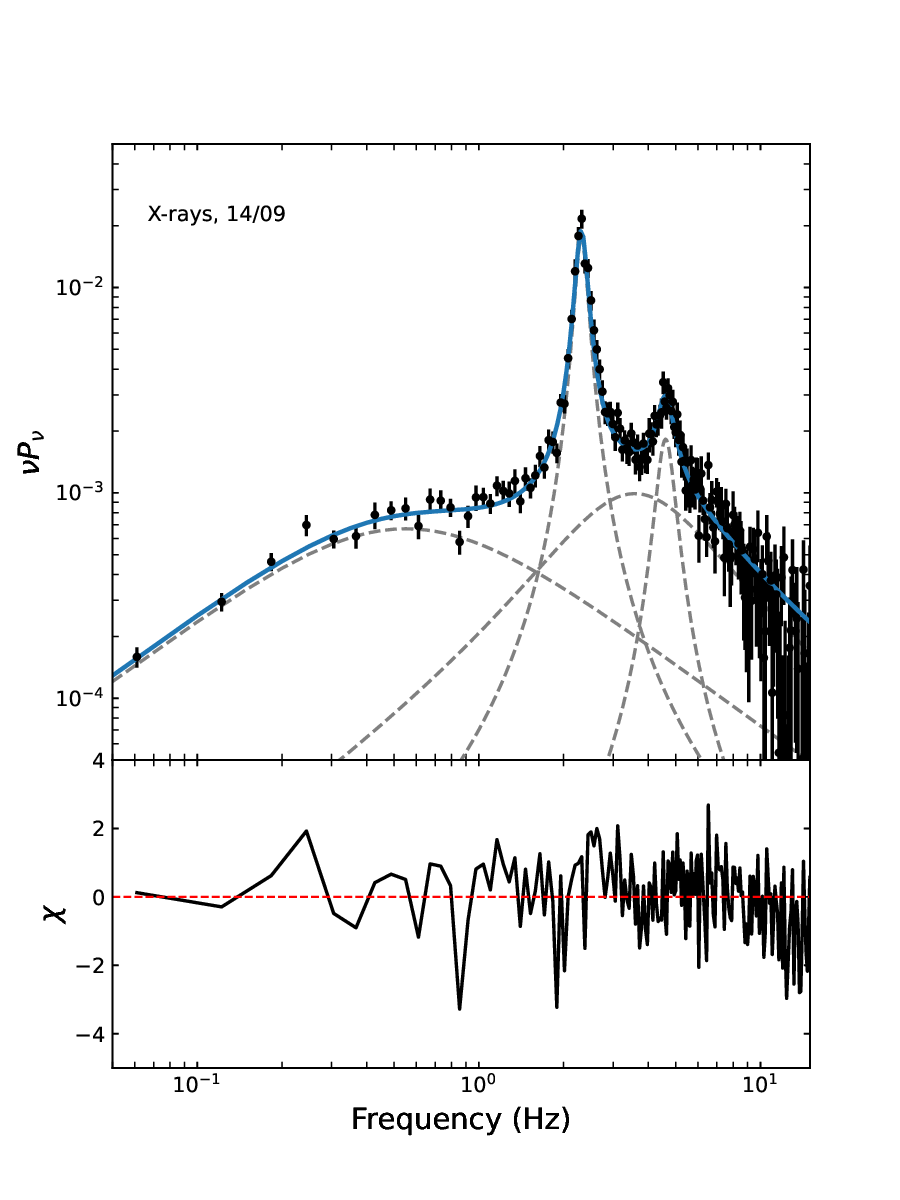}
  \caption{ \textit{Top Panels:} PDSs fit for the two \textit{NICER} epochs. Grey dashed lines represent the individual Lorentzian components, while blue continuous lines show their sum. \textit{Bottom Panels:} Residuals computed as data-model/ $\sigma$. }
  \label{fig:nicer_res}
\end{figure*}

\begin{figure*}
  \centering
    \includegraphics[scale=0.5]{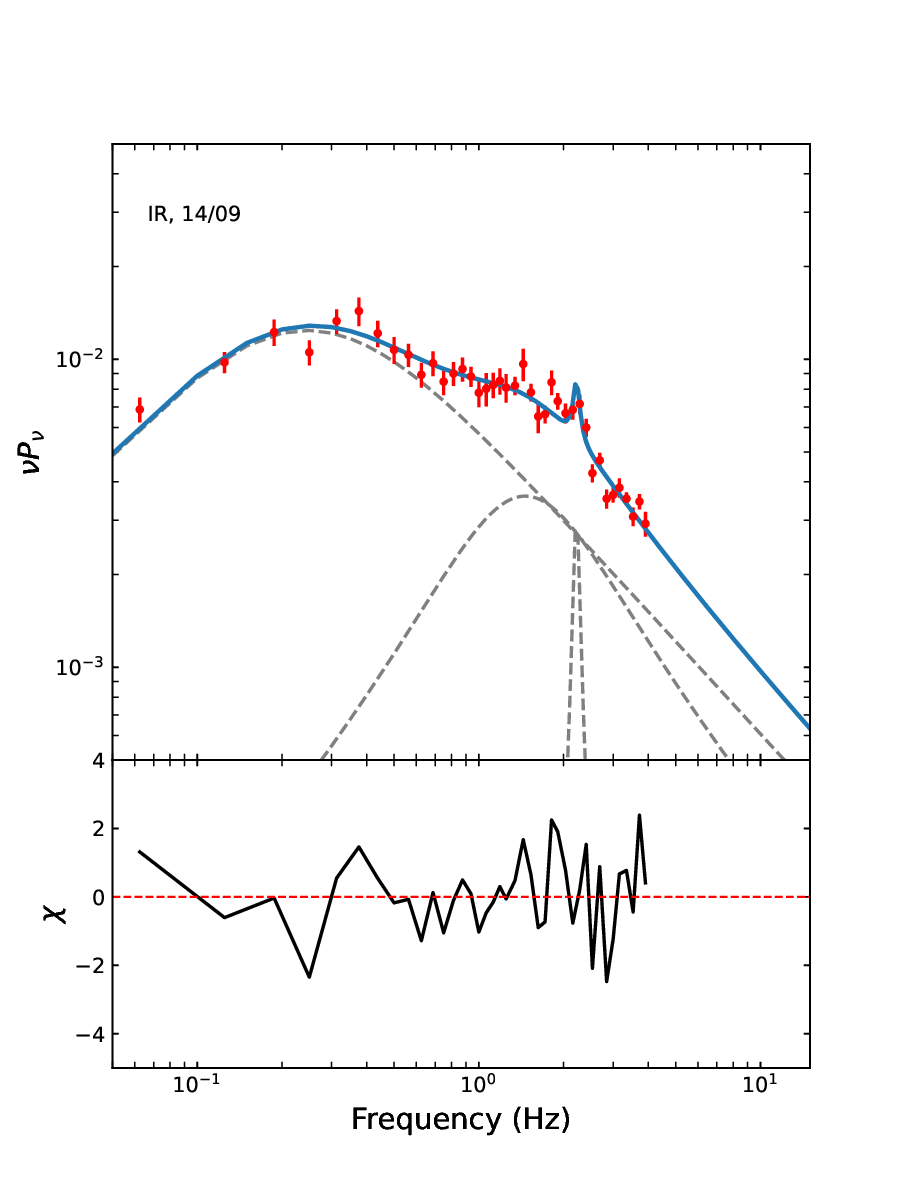}
  \includegraphics[scale=0.5]{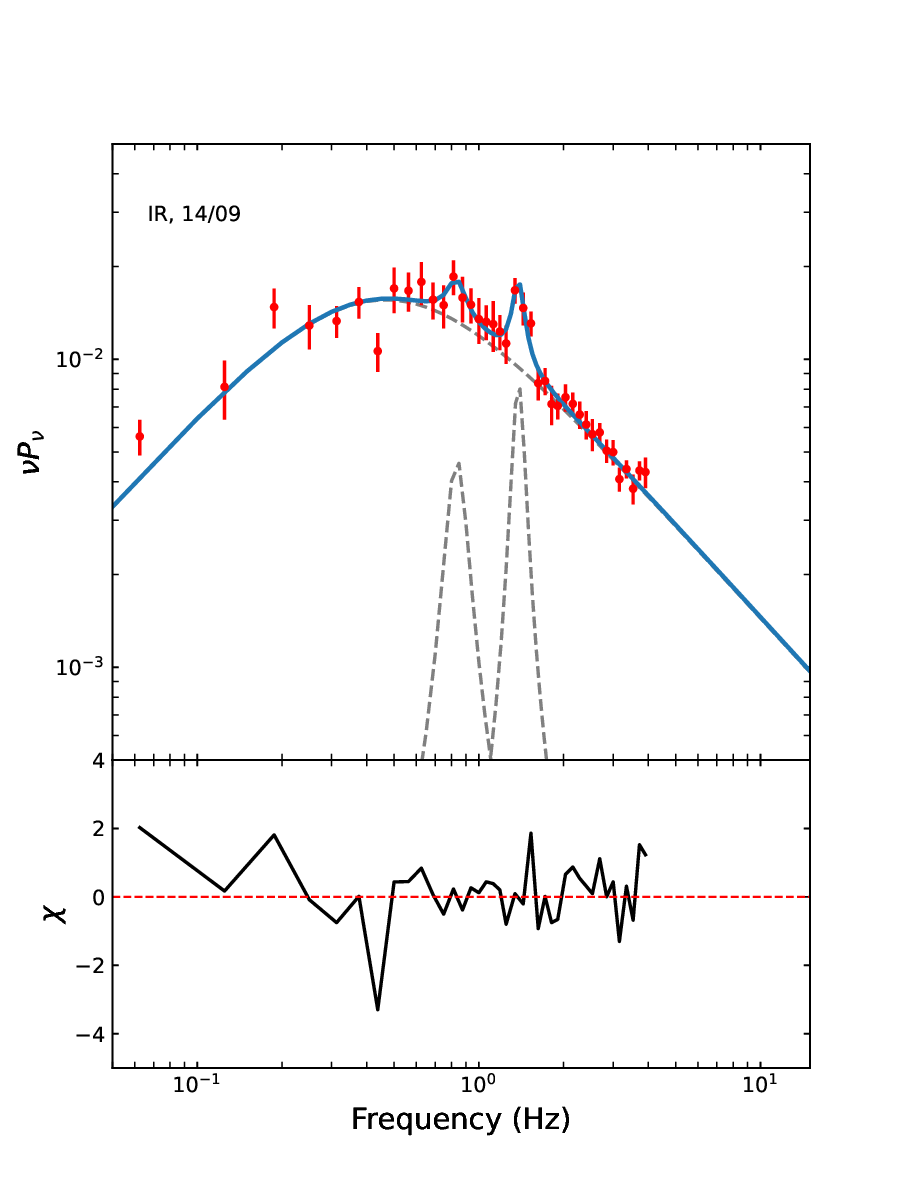}
  \caption{ PDSs fit and residuals for the two \textit{HAWK-I} epochs. Colours of the fitting models and structure of the panels are the same of Fig. \ref{fig:nicer_res}. }
  \label{fig:hawki_res}
\end{figure*}

\begin{figure*}
  \centering
  \includegraphics[scale=0.5]{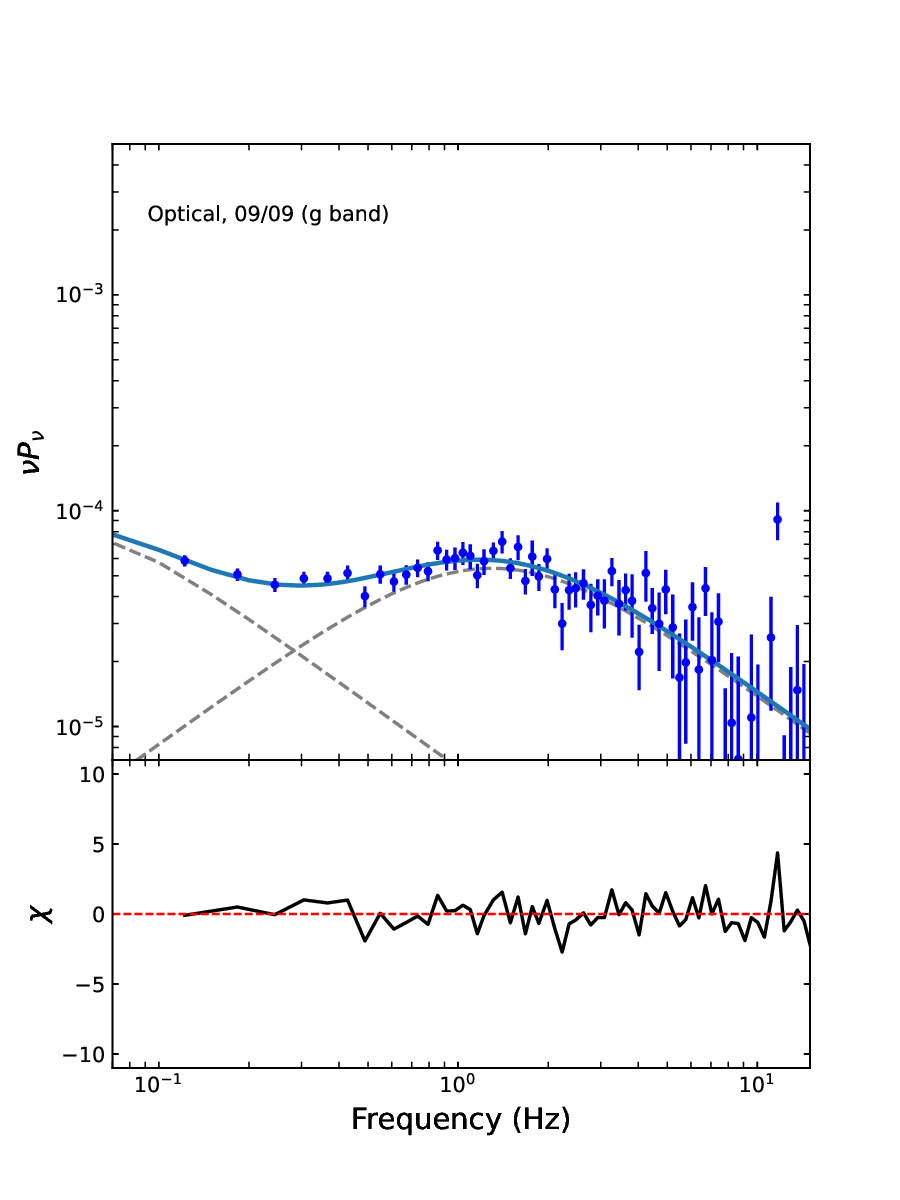}
    \includegraphics[scale=0.5]{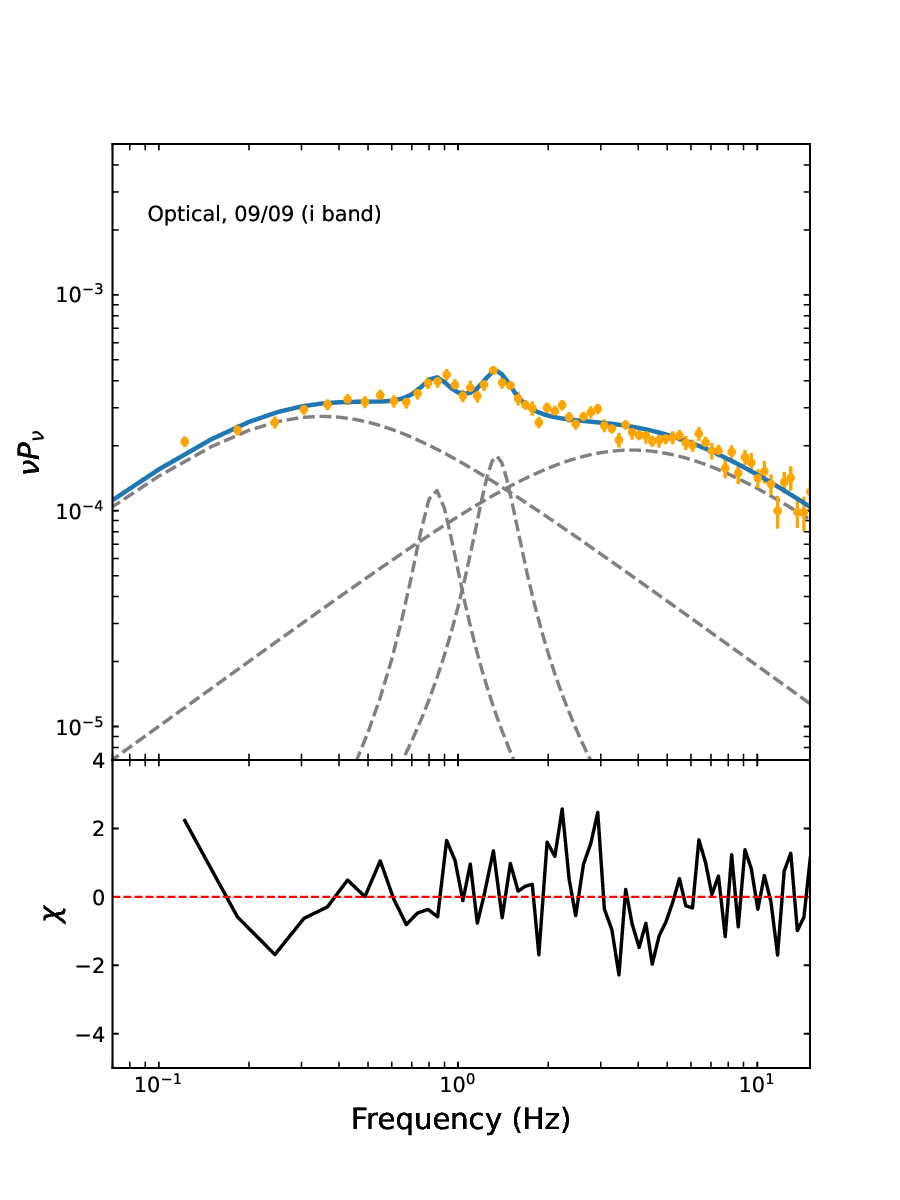}

  \caption{ PDSs fit and residuals for the  PDSs fit and residuals for the two filters in the \textit{ULTRACAM} epoch. Colours of the fitting models and structure of the panels are the same of Fig. \ref{fig:nicer_res}. }
  \label{fig:ucm_res}
\end{figure*}

\begin{figure*}
  \centering
  \includegraphics[scale=0.5]{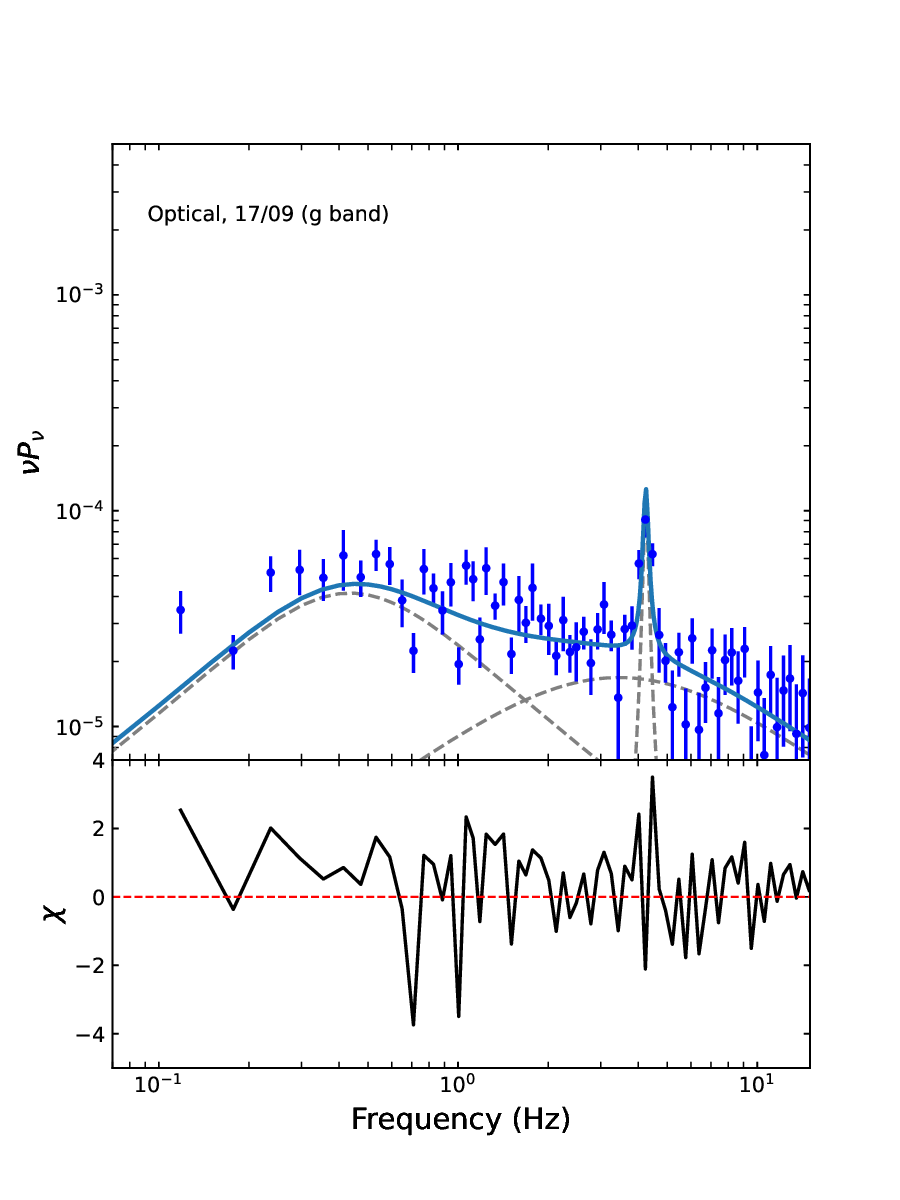}
    \includegraphics[scale=0.5]{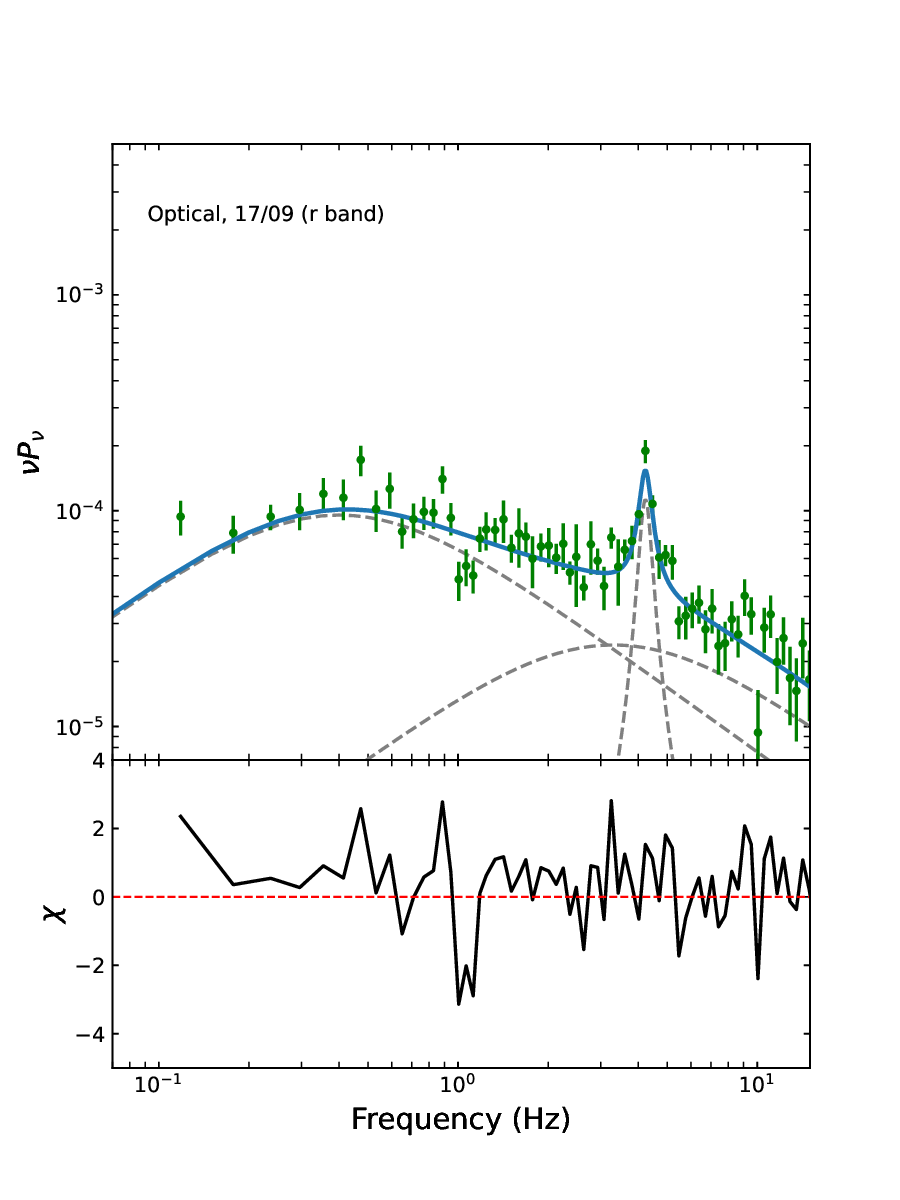}

  \includegraphics[scale=0.5]{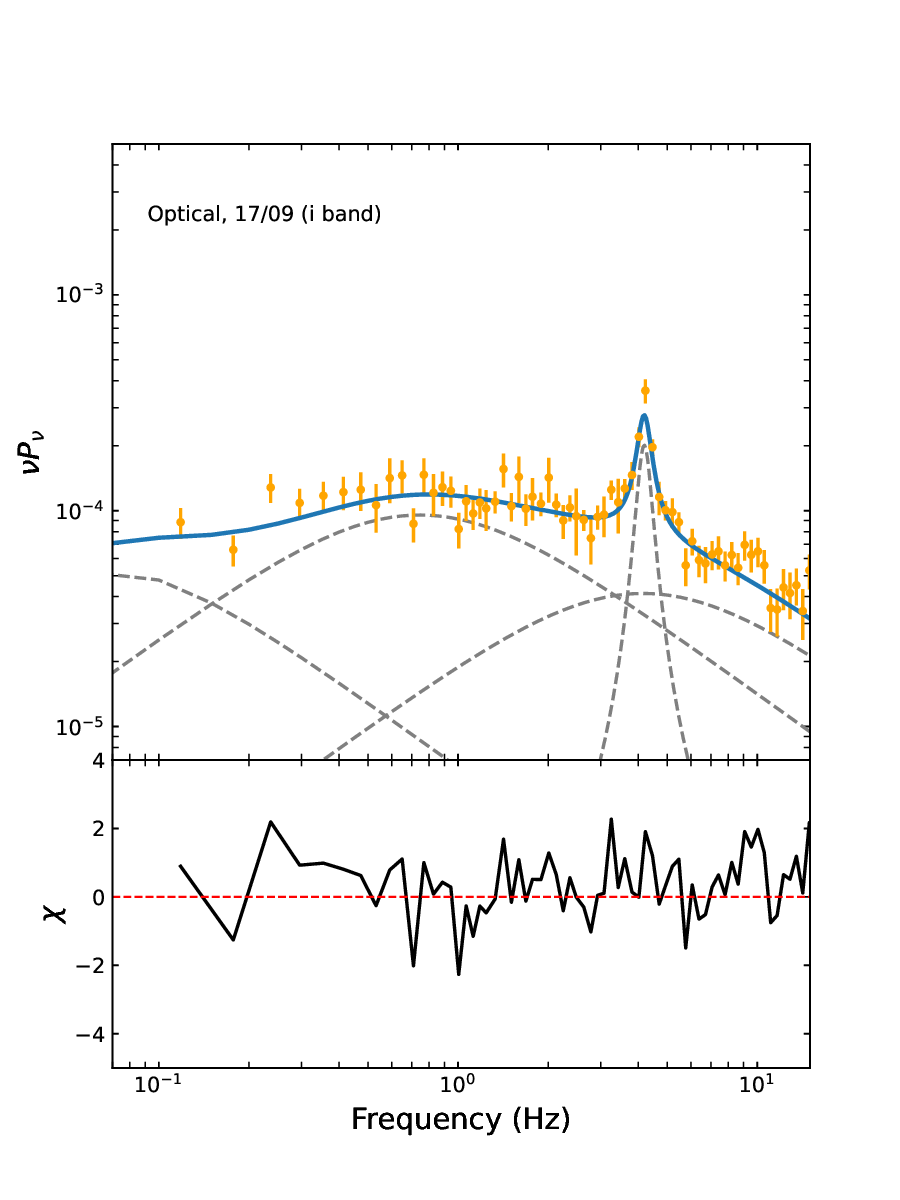}
    \includegraphics[scale=0.5]{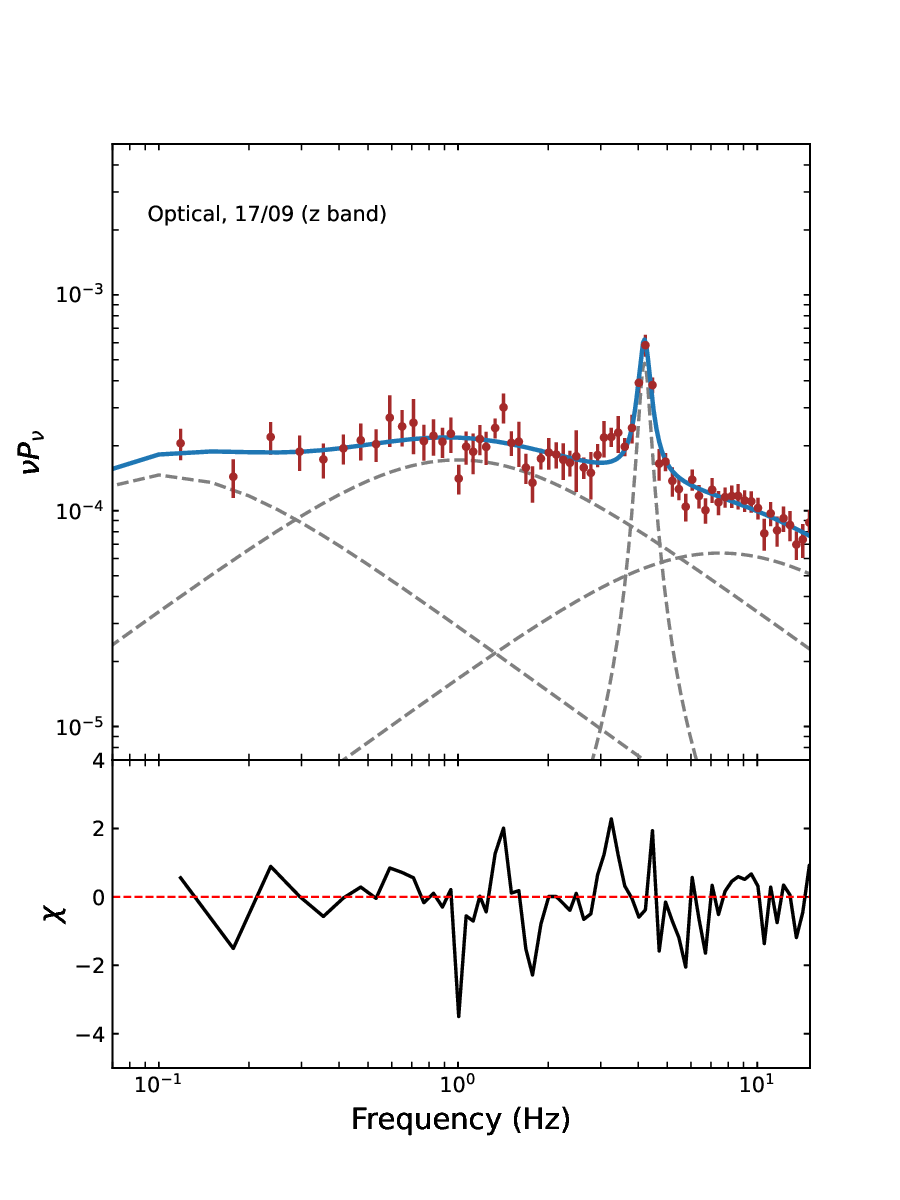}

  \caption{ PDSs fit and residuals for the PDSs fit and residuals for the four filters in the \textit{HiPERCAM} epoch. Colours of the fitting models and structure of the panels are the same of Fig. \ref{fig:nicer_res}. }
  \label{fig:hpcm_res}
\end{figure*}

\bsp	
\label{lastpage}
\end{document}